\documentclass[12pt]{article}
\pdfoutput=1
\usepackage{amssymb}
\usepackage{amsmath}
\usepackage{graphicx}
\usepackage{hyperref}
\usepackage{bm}
\usepackage{color}

\ifx\affiliation\undefined %Then we are using plain Latex class
\def\affiliation#1{\date{\normalsize #1\\ \today}}
\usepackage[margin=1in]{geometry}
\usepackage[sort&compress,numbers]{natbib}
\else % We are using jpp class.
\fi

\def\ket#1{|#1\rangle}
\def\bra#1{\langle#1}
\def\sech{{\,\rm sech}}
\def\etothe#1{{\rm e}^{#1}}
\def\a{{\bf a}}
\def\M{{\bf M}}
\begin{document}
\title{Multimode theory of electron hole transverse instability}
\author{Xiang Chen and I H Hutchinson}
\affiliation{Plasma Science and Fusion Center,\\ Massachusetts
  Institute of Technology, Cambridge, Massachusetts, USA} 

\ifx\altaffiliation\undefined\maketitle\fi % Not JPP

\begin{abstract}
We present Vlasov-Poisson 3-D linear stability analysis of an
initially planar electron hole structure, solving for the distribution
function by integration along unperturbed orbits. The non-sinusoidal
potential perturbation shape (parallel to $B$) is expanded in
eigenfunctions of the adiabatic Poisson operator, generalizing the
prior assumption of a rigid shift of the equilibrium.  We show that
the shiftmode is then modified by a second discrete mode plus an
integral over a continuum of wave-like modes. A rigorous treatment
shows that the continuum can be approximated effectively by a single
mode that satisfies the external wave dispersion relation, thus making
the perturbation a weighted sum of three modes. We find numerically
the solution for the complex instability frequency, and the
corresponding three mode amplitudes determining the perturbation
eigenmode. This multimode analysis refines the accuracy of the prior
single mode results, giving slightly higher growth rates at most
parameters, as expected from the extra mode shape freedom. Oscillating
modes near stability boundaries have larger mode distortions which
help explain PIC simulations that observe instability up to $\sim20$\%
beyond the prior shiftmode thresholds, and narrowing of the
perturbation. At high magnetic field, the multimode analysis predicts
a reduction of the already small growth rate.
\end{abstract}

\section{Introduction}

Electron holes are nonlinear solitary BGK \citep{Bernstein1957}
electrostatic structures sustained by electron trapping
\citep{Hutchinson2017}, which occur widely in space plasmas For a
recent summary of observations see e.g. \citep{Lotekar2020} and
references therein.  The theoretical stability of electron holes to
motions with sinusoidal variation transverse to the magnetic field has
recently been extensively analyzed assuming that the relevant unstable
perturbation is a parallel, kink-like, uniform shift of the hole
position.
\citep{Hutchinson2018,Hutchinson2018a,Hutchinson2019,Hutchinson2019a}. In
that work the dispersion relation's complex frequency $\omega$, has
been found using a ``Rayleigh Quotient'' \citep{Parlett1974} which
provides an approximation accurate to second order in any deviations
from the exact unstable mode structure, and is equivalent to solving
the hole's overall momentum balance \citep{Hutchinson2016}. The
results are in reasonable agreement with particle in cell (PIC)
simulations, but some discrepancies have been noted. The purpose of
the present work is to discover whether more accurate mode parallel
shape determination, including deviations from the shiftmode, can
explain those discrepancies, and to quantify by analysis how important
the deviations are.

Two main phenomena have been observed in transverse instability PIC
simulations that are not well represented by the shiftmode
analysis. They are (1) narrowing of the unstable mode structure
relative to the shiftmode, when near marginal oscillatory
stability
\citep{Hutchinson2019}; (2) generation of long parallel length
external ``streaks'' or waves on the whistler branch, during
oscillatory instability at high magnetic field
\citep{Hutchinson2019a}.

Our approach to managing a generalization of the unstable mode of a
Vlasov-Poisson problem, following longstanding mathematical
analysis
\citep{Lewis1979,Symon1982}, is to represent the perturbation of
potential in terms of the orthogonal eigenmodes of a judiciously
chosen differential operator. In the present context, the operator
generally used
\citep{Lewis1982} represents the Poisson equation for
steady (or very slowly varying) potential ($\nabla^2\phi-n=0$ in
appropriately normalized units). The charged particle density
difference $n_0(z)$ associated with the potential equilibrium
$\phi_0(z)$, can be determined from this equation. Then for
infinitessimally slow linearized potential perturbations $\phi_1$
about this equilibrium the perturbed density is
$n_1=\phi_1dn_0/d\phi_0$, and the resulting Poisson equation can be
written using the operator $V_a \equiv(\nabla^2-dn_0/d\phi_0)$ acting
on $\phi_1$ as $V_a\phi_1=0$. This $V_a$ is called the ``adiabatic''
operator, and the associated density perturbation $n_1$ the adiabatic
density (perturbation). For low frequencies, expansion of the
potential perturbation is most naturally in terms of the eigenmodes of
$V_a$. For purely growing instabilities $\omega=0+i\omega_i$, at the
threshold $\omega_i\sim 0$, evidently the adiabatic response is nearly
equal to the total because changes are infinitessimally slow; so the
non-adiabatic part $\tilde V$ of the operator is small compared with
$V_a$ in the perturbed Poisson equation. Consequently to lowest order
the perturbation unstable mode is equal to the eigenmode of $V_a$ with
zero eigenvalue. For a solitary potential structure in a uniform
background, such a zero eigenvalue always exists, and its eigenmode
has the form of a uniform shift of the equilibrium.

By the preceding argument, determining where $\omega_i=0$, i.e. the
stability threshold, can be accomplished exactly using just the
shiftmode, provided that the real part $\omega_r$ of the mode
frequency is zero. However, some hole instabilities are oscillatory:
$\omega_r\not=0$, $\omega_i>0$ and even if they are not we may wish to
find a finite $\omega_i$ value. Then the unstable mode is not purely
the shiftmode, and the extent to which it includes contributions from
other eigenmodes of $V_a$ becomes an important question. It can be
explored by carrying the perturbation analysis to first order in the
other eigenmodes, in much the way that time-independent perturbation
theory is used in quantum mechanics. This approach was successfully
pursued in an early study of the one-dimensional instability of a
train of electron holes that leads eventually to hole merger
\citep{Schwarzmeier1979}. However, the few subsequent efforts to apply
it \citep{Schamel1982,Schamel1987,Collantes1988} have been of limited
utility either because of expanding about the wrong eigenmode
(symmetric and not having zero eigenvalue), or, more fundamentally
because of adopting inappropriate approximations to the solution of
the Vlasov equation \citep{Jovanovic2002}, which constitutes the
complementary (and more difficult) part of the Poisson-Vlasov system:
the non-adiabatic perturbation. A more recent one-dimensional analysis
\citep{Dokgo2016} of $\sech^2(z/\ell)$ shaped holes considered their
sole antisymmetric discrete adiabatic eigenmode, which is simply the
shiftmode.

A separate thread of this question is the coupling, observed in
simulations\citep{Oppenheim1999,Oppenheim2001b,Lu2008}, of hole
instabilities at high magnetic field to long-wavelength external
perturbations ``streaks'' identified as belonging to the cold-plasma
whistler branch. Previous analyses
\citep{Newman2001a,Vetoulis2001,Berthomier2002} theorized that
coupling to these waves was the primary cause of instability at high
field, but adopted unjustified ad hoc expressions for the
coupling. Recently it was shown that the high-field instability is not
explicitly caused by the coupling \citep{Hutchinson2019a}, but since
simulations show external coupling can affect the instability, an
interesting question remains how to calculate it self-consistently.

The present approach solves the Vlasov problem without expansion, by
integration over the prior orbit, numerically in the hole region.  In
the process the solution for wave-coupling emerges rigorously from the
mathematical analysis. Moreover we avoid any perturbative
approximation for the relative amplitudes of the different modes.
Instead we show how, in addition to a single extra discrete eigenmode
contribution, the eigenmode continuum contribution can be
approximately represented through a single amplitude corresponding to
the external wave dispersion relation.  Section 2 explains the
eigenmodes and the expansion. Section 3 shows for the Vlasov operator
form how the continuum modes can be included and reduced to
effectively a single contribution. Section 4 discusses the numerical
and analytic evaluations needed. Section 5 presents results.

\section{Eigenmode Expansion}
Let us adopt a minimalist bra-ket notation for the adiabatic
eigenmodes in which we expand: $\ket{e}$, where the label $e$ being
either real $p,q,\dots$, or integer, $j,l,\dots$, will denote
respectively continuum or discrete eigenmodes. The inner product of
any two Hilbert-space vectors (complex potential functions of $z$)
denotes an overlap integral over (parallel) spatial coordinate
$\bra{u}\ket{s}\equiv \int u^*(z)s(z)dz$. Insofar as the eigenmodes
are orthonormal, we write $\bra{u}\ket{s}=\delta_{us}$, where for
continuum modes such that $u$ and $s$ are real parameters this is
(approximately) a Dirac delta function, $\delta_{us}=\delta(u-s)$,
whereas for discrete modes $\delta_{jl}$ is the Kronecker delta.

\subsection{Adiabatic Response Eigenmodes}
In units scaled to Debye length, electron
background density, and electron thermal energy, the one-dimensional
Poisson equation, assuming immobile uniform ion density, is
$d^2\phi/dz^2=n-1$. The equilibrium analyzed is chosen to be of the
form $\phi_0=\psi\sech^4(z/4)$. Writing for brevity $S=\sech(z/4)$ and
$T=\tanh(z/4)$ and denoting the $z$-derivative by prime, the
linearized equilibrium (``adiabatic'') Poisson operator for this hole
potential can then be written
\begin{equation}
  \label{poissonop}
  V_a = \left[{d^2\over dz^2} - {dn\over d\phi_0}\right]
  = \left[{d^2\over dz^2} - {\phi_0^{'''}\over \phi_0'}\right]
  = \left[{d^2\over dz^2} +{30\over 16}S^2- 1\right].
\end{equation}
Notation $V$ reminds us that this is partly the Vlasov operator,
transforming a potential into a density, and the subscript $a$ denotes
adiabatic (meaning steady).  The eigenmodes of this
operator, satisfying $V_a\ket{u}=\lambda \ket{u}$, can be found (see
appendix A) by
applying the raising operator $4 {d\over dz}-l T$ for $l=1,2,3,4,5$ to
the function $ \etothe{uz/4}$ yielding
\begin{equation}
  \label{ketu}
  \begin{split}
 \ket{u}(z)= &\exp(uz/4)\{[-15u^4 + (420S^2 - 225)u^2 - 945S^4 +
 840S^2 - 120]T \\
 &\qquad\qquad+ u[u^4 + (-105S^2 + 85)u^2 + (945S^4 - 1155S^2
 +274)]\}.
\end{split}
\end{equation}
For real $u$, as $u z\to+\infty$ the mode is bounded (and tends to
zero) only if $u$ is one of the discrete roots of the polynomial in
the braces obtained by letting $S\to 0$ and $T\to {\rm sign}(u)$. These
are $u=j=1,2,3,4,5$. The odd numbered discrete modes are symmetric in
$z$, and the even numbered are antisymmetric. In contrast, imaginary
$u$-values $u=ip$ give rise to the continuum modes which are formally
finite at infinity. Their overlap integral over a finite domain
exists; and, as the domain tends to infinity, it tends to a delta
function. The continuum eigenmodes have definite parity only if $u$
reverses sign with $z$; so we write $u=i \sigma_z p$ where
$\sigma_z={\rm sign}(z)$, and positive $p$ represents antisymmetric
outwardly-propagating waves.

The corresponding eigenvalues are
\begin{equation}
  \label{eigenvalues}
  \lambda_u= u^2/16-1=-p^2/16-1.
\end{equation}
Normalized so that
$\bra{j}\ket{l}=\delta_{jl}$ they are given in Table \ref{discrete}. 
\begin{table}
  \center
\begin{tabular}{ccc}\
  Mode & Eigenvalue & Normalized Form\\
  $\ket{1}$& -15/16&$S(21S^4 - 28S^2 + 8)\sqrt{30}/32$\\
  $\ket{2}$& -12/16&$TS^2(3S^2 - 2)\sqrt{105}/8$\\
  $\ket{3}$& -7/16 &$-S^3(9S^2 - 8)\sqrt{105}/32 $\\
  $\ket{4}$&  0 &$-3TS^4\sqrt{70}/16 $\\
  $\ket{5}$&  9/16 &$3S^5\sqrt{35}/32 $\\
\end{tabular}
\caption{Discrete eigenmodes.\label{discrete}}
\end{table}
\noindent
In particular $\lambda_4=0$ for the shiftmode
$\ket{4}\propto d\phi_0/dz$, and it is the predominant perturbation in
essentially all linear electron hole instabilities. Moreover, it
couples only to antisymmetric modes; so $\ket{2}$ is the only other
discrete mode that needs to be considered.

\begin{figure}\center
  \includegraphics[width=0.6\hsize]{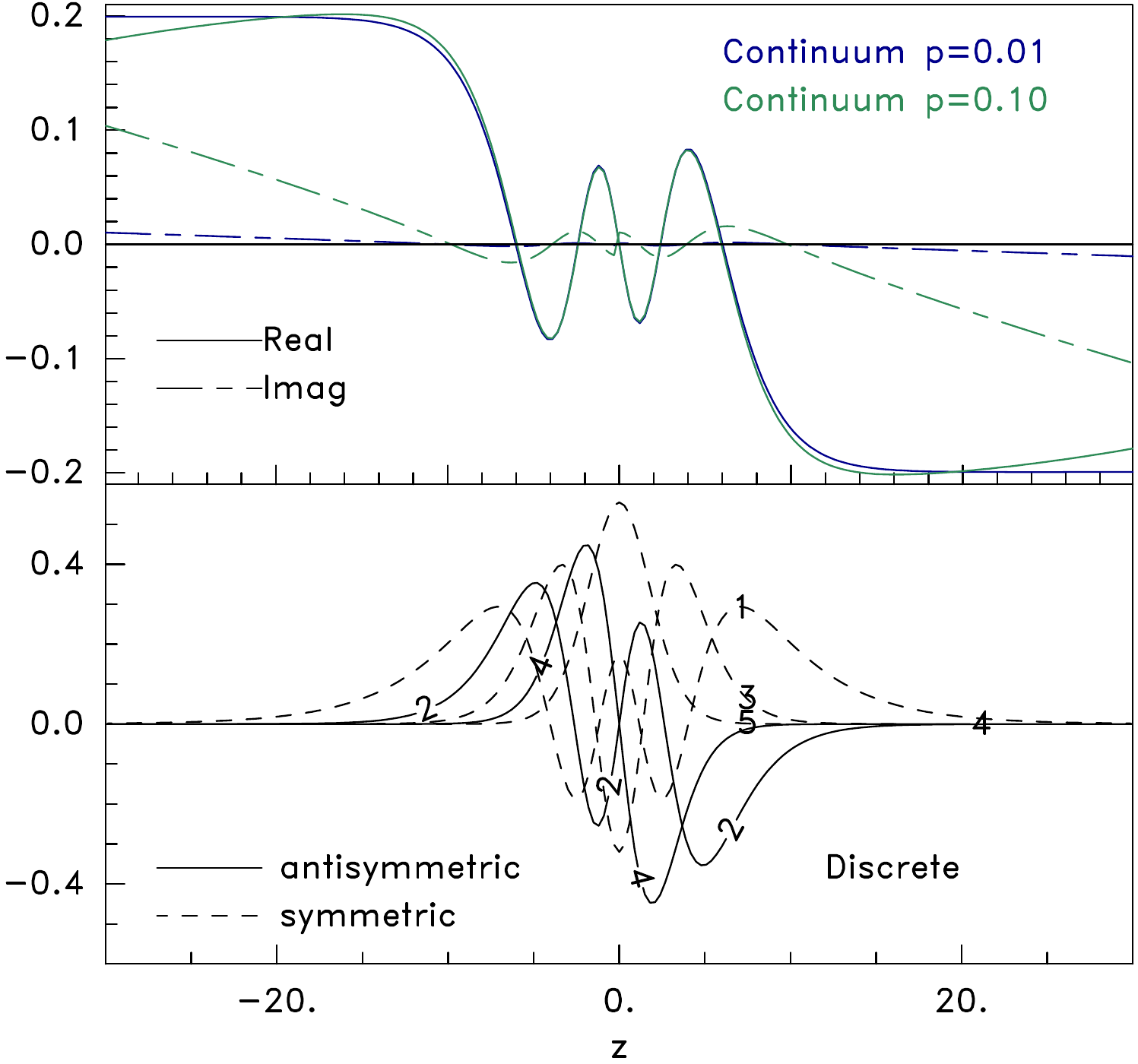}
  \caption{Eigenmodes of the adiabatic response operator.\label{modeplots}}
\end{figure}

It can
be shown that the the continuum modes are ``normalized'', in the sense
that then $\bra{p}\ket{q}=\delta_{pq}=\delta(p-q)$, by dividing eq.\
(\ref{ketu}) by the factor
\begin{equation}
  [8\pi(p^2+1^2)(p^2+2^2)(p^2+3^2)(p^2+4^2)(p^2+5^2)]^{1/2}.
  \label{normfactor}
\end{equation}
Far
from the hole ($|z|\gg1$) the normalized oscillatory continuum modes are
sinusoidal with amplitude $1/\sqrt{8\pi}=0.19947$, and parallel wavenumber
$|k_\parallel|=p/4$. The eigenmodes are plotted in Fig.\
\ref{modeplots}.

\subsection{Including Non-adiabatic Response}
We now must consider the full linearized Poisson equation including the
non-adiabatic response arising from the solution to the time-dependent
Vlasov equation, $\tilde{V}$, as well as the adiabatic response
$V_a$. The form of the non-adiabatic Vlasov operator $\tilde{V}$ will
be discussed later. For a perturbation
$\phi_1(y,z,t)=\hat\phi(z){\rm e}^{i(k_\perp y-\omega t)}$ with
transverse wavenumber $k_\perp$ and frequency $\omega$, Poisson's
equation becomes
\begin{equation}
  \label{fullpois}
  (-k_\perp^2+V_a+\tilde{V})\ket{\hat\phi}=0,
\end{equation}
in which it is convenient to regard $k_\perp^2\equiv\lambda_\perp$ as
the full eigenvalue.  We suppose the solution for potential can be
expanded as a sum of scalar amplitudes $a_u$ times the eigenmodes of
$V_a$, $\ket{u}$:
\begin{equation}
  \label{ketphi}
  \ket{\hat\phi}=\sum_u a_u \ket{u},
\end{equation}
where the summation notation also includes an integral over the
continuum modes.

Then we can invoke the orthogonal properties of the
adiabatic eigenmodes and form the inner product
\begin{equation}
  \label{VaVt}
  \bra{s}|(%-16k_\perp^2+
  V_a+\tilde{V})\ket{\hat\phi}
  = %-16k_\perp^2+
  \lambda_s\bra{s}\ket{s}a_s+\sum_u\bra{s}|\tilde{V}\ket{u}a_u,
\end{equation}
which must be equal to $\lambda_\perp\bra{s}\ket{s}a_s$ to satisfy eq.\ \ref{fullpois}.
In particular, choosing the predominant mode $\bra{4}|$ for
$\bra{s}|$, for which $\lambda_4=0$, we get an eigenvalue equation
$0=(\lambda_4-\lambda_\perp)\bra{4}\ket{4}a_4+\sum_u\bra{4}|\tilde{V}\ket{u}a_u$.

If all the amplitudes $a_u$ for $u\not=4$ are negligible, then
$\tilde{V}$ contributes a correction ($\lambda_{4}^{(1)}$) to the eigenvalue
\begin{equation}
  \label{lam41}
  \lambda_{4}^{(1)}= {\bra{4}|\tilde{V}\ket{4}\over\bra{4}\ket{4}}=\lambda_\perp-\lambda_4
\end{equation}
The expression $\bra{4}|\tilde{V}\ket{4}/\bra{4}\ket{4}$ is the ``Rayleigh
quotient'' approximation for the eigenvalue of $V_a+\tilde{V}$
(because $\lambda_4=0$). It is
physically the (normalized) jetting force on particles because of the
unperturbed electric field $-d\phi_0/dz\propto\bra{4}|$, acting on the
non-adiabatic density perturbation $\tilde n = \tilde{V}\ket{4}$, integrated
over the entire hole. It balances the (normalized) Maxwell shear
stress from the transverse kinking of the hole ($k_\perp^2$) to
make the total force zero.

Actually $\tilde{V}$ is a complicated nonlinear function of the
complex frequency $\omega$ of the mode; and for specified $k_\perp$
the dispersion relation between $\omega$ and $k_\perp$ must be solved
by some kind of iterative procedure searching for an $\omega$ that
satisfies eq.\ (\ref{lam41}). The imaginary part of $\omega$ thus found
determines the stability of the hole. This approximation has yielded
stability results
\citep{Hutchinson2018a,Hutchinson2019,Hutchinson2019a} that are in reasonable (but not perfect) agreement
with simulation.  The question at hand is whether analysis can
determine approximately the magnitude of the other coefficients $a_u$
for $u\not=4$, and therefore give a more accurate perturbation
structure $\ket{\hat\phi}$ and $\omega$.

If instead of the approximation (\ref{lam41}), we are able to evaluate all the
matrix coefficients $\bra{s}|\tilde{V}\ket{u}$ then in principle we can regard eq.\
(\ref{VaVt}) instead of (\ref{lam41}), as a matrix eigen-system we must
solve to find the $\omega$ that permits a non-zero solution for the
vector $a_u$. The off-diagonal matrix entries $s\not=u$ are the
coupling of the potential modes by the non-adiabatic Vlasov operator
$\tilde{V}$.  The condition for the existence of a solution is that the
determinant of the matrix
$[(-\lambda_\perp+\lambda_s)\bra{s}\ket{s}\delta_{su}+\bra{s}|\tilde{V}\ket{u}]$ should
be zero.

Such a program faces formidable practical challenges, however, because
each evaluation of $\tilde{V}\ket{u}$, requires a computation
involving multiple-dimension integrations over space and velocity
distribution --- repeated for each mode $\ket{u}$ and each adjustment
of $\omega$.  Moreover, in principle, the continuum contains
infinitely many modes, and the matrix contains the square of the
number of modes.  Obviously we require this number to be reduced. We
can immediately restrict our attention just to antisymmetric modes, since by
assumed symmetry no coupling to symmetric modes occurs. We later show
how the entire continuum contribution can be reduced to a single
amplitude. Continuum modes $\ket{p}$ extend to
$|z|=\infty$: an integration range that for computation needs to be
reduced, and have formally divergent inner products. We will show how
these difficulties are negotiated.

Formally, any mode for which $\tilde{V}\ket{u}a_u$ is not negligible
must be retained in the sum of $\bra{s}|\tilde{V}\ket{u}$ in eq.\
\ref{VaVt}, giving for the dominant mode
\begin{equation}
  \label{yes4}
  0=(\lambda_4-\lambda_\perp)\bra{4}\ket{4}a_4+\bra{4}|\tilde{V}\ket{4}a_4+\sum_{u\not=4}\bra{4}|\tilde{V}\ket{u}a_u,
\end{equation}
but also for $s\not=4$
\begin{equation}
  \label{not4}
 0=(\lambda_s-\lambda_\perp)\bra{s}\ket{s}a_s+\bra{s}|\tilde{V}\ket{4}a_4+\sum_{u\not=4}\bra{s}|\tilde{V}\ket{u}a_s.
\end{equation}

The well known approach of time-independent perturbation theory in
elementary quantum mechanics
\citep[see e.g.][Section 43]{Dirac1958}
regards $\tilde{V}$ as systematically small, and takes $a_s$ for
$s\not=4$ also to be first order small relative to $a_4$. The first
order approximation of eq.\ \ref{not4} then drops the final sum term,
giving
\begin{equation}
  \label{assim}
  a_s \simeq {\bra{s}|\tilde{V}\ket{4}a_4\over (\lambda_\perp-\lambda_s)\bra{s}\ket{s}}.
\end{equation}
Substituting back (with $s\to u$) into eq.\ \ref{yes4}, the
eigenvalue to second order is
\begin{equation}
  \label{eval2}
 \lambda_\perp\simeq \lambda_4+\lambda_{4}^{(1)}+\lambda_{4}^{(2)}=\lambda_4
  +{\bra{4}|\tilde{V}\ket{4}\over \bra{4}\ket{4}}
    +\sum_{u\not=4}
    {\bra{4}|\tilde{V}\ket{u}\bra{u}|\tilde{V}\ket{4}\over
      (\lambda_\perp-\lambda_u)\bra{4}\ket{4}\bra{u}\ket{u}}.
\end{equation}
Normally the substitution $\lambda_\perp\simeq\lambda_4$ is made in
the final sum; but we do not need to do that since we consider
$\lambda_\perp$ to be given and $\omega$ to be changed to achieve
equality in this equation.

These perturbation equations (\ref{assim}) and (\ref{eval2}) are
appropriate for coupled discrete modes $\ket{j}$ when the amplitudes
$a_u\bra{u}\ket{u}$ for $u\not=4$ are small compared with
$a_4\bra{4}\ket{4}$. Those equations give the first order mode
amplitude and second order eigenvalue correction from them.  However,
in our case, for the continuum modes, $\tilde{V}$ is not
systematically small everywhere, and an altered approach appropriate
to the actual form of $\tilde{V}$ must be adopted. We shall see that
in order to obtain converged integrals for the relevant continuum
inner products it is necessary to use the difference between
$\tilde V \ket{q}$ and a pure wave operator expression
$\tilde V_w\ket{q}$.

\section{The Eigenmode Coefficients}
\subsection{The Non-adiabatic Linearized Vlasov Operator}

The operator $\tilde{V}$ transforms a potential perturbation $\ket{u}$
into a non-adiabatic density perturbation $\tilde n$, both of which
are complex functions of $z$. It does so by solving the linearized
time-dependent Vlasov equation for the non-adiabatic distribution
function perturbation $\tilde f(z,v)$ and integrating it:
$\tilde n =\int \tilde f dv$. The solution can be found in terms
of an integral over past time along the linearized Vlasov equation's
``characteristic'', that is, the unperturbed orbit $\bm x(t)$, giving
an integral expression
\begin{equation}
  \label{eq:phim}
  \Phi(\bm x,\bm v,t)\equiv 
  \int_{-\infty}^t \phi_1(\bm x(\tau),t-\tau ) d\tau,
\end{equation}
where $\phi_1(\bm x,t)$ is the potential perturbation and
$\bm x(\tau)$ is the past position of the unperturbed orbit that has
velocity $\bm v$ at $(\bm x,t)$. The units of time adopted are
$1/\omega_{pe}$: the inverse of the electron plasma frequency. When a
uniform magnetic field in the $z$-direction is present with electron
cyclotron frequency $\Omega$ and the background
perpendicular velocity distribution is Maxwellian, then the
transverse perpendicular velocity dependence can be expressed as a sum
over cyclotron harmonics
\citep{Hutchinson2018a}.  The non-adiabatic
parallel distribution function perturbation is then
\begin{equation}\label{eq:ftmagnetic}
    \tilde f(z,y,v,t) = {\rm e}^{i(k_\perp y-\omega t)}
    \sum_{m=-\infty}^\infty \tilde f_m(z,v)= {\rm e}^{i(k_\perp
      y-\omega t)} \sum_{m=-\infty}^\infty b_m \Phi_m,
\end{equation}
with
\begin{equation}
  \label{mweight}
   b_m=i\left[\omega_m
  {\partial f_{\parallel0}\over \partial W_\parallel}
  +m\Omega {f_{\parallel0}\over T_\perp}\right]
q_e{\rm e}^{-\zeta_t^2}I_m(\zeta_t^2),\qquad
\Phi_m=\int_{-\infty}^t \hat\phi(z(\tau))\etothe{-i\omega_m(\tau-t)}d\tau.
\end{equation}
Here $\zeta_t^2=k^2T_\perp/\Omega^2m_e$, where $T_\perp$ is the
perpendicular temperature, $I_m$ is the modified Bessel function, and
$\omega_m=m\Omega+\omega$. The unperturbed equilibrium parallel
distribution function depends only on parallel energy $W_\parallel$
(not $z$ directly) and is written $f_{\parallel 0}$, but for the
perturbed parallel distributions we omit the $\parallel$ subscript for
brevity, and from now on omit the $y,t$ dependencies as implicitly
$ {\rm e}^{i(k_\perp y-\omega t)}$ (so in $\Phi_m$, the upper limit is
$t=0$). The prior integral $\Phi_m$ is a function of parallel position
$z$ and velocity $v$. The weight $b_m$ is independent of $z$ but
depends on $W_\parallel$ through the $f_{\parallel0}$ distribution.
We can regard each harmonic $m$ as giving a perturbed density
contribution $\tilde n_m(z)=\int \tilde f_m dv$. When $\hat\phi$ is a
sum over modes $\ket{u}$,
\begin{equation}
  \label{nmexpansion}
\tilde  n_m=\tilde{V}_{m}\ket{\hat\phi}=\tilde{V}_{m}\sum_u a_u\ket{u}=\int \sum_u a_u
  \tilde f_{um} dv=\sum_u a_u \tilde n_{um},
\end{equation}
where $\tilde f_{um}$ denotes $\tilde f_m$ with
$\ket{\hat\phi}=\ket{u}$ substituted in eq. (\ref{eq:ftmagnetic}), and
similarly $\tilde n_{um}=\int \tilde f_{um}dv$. Then
$\tilde V \ket{\hat\phi}=\sum_m\tilde
V_m\ket{\hat\phi}=\sum_{u,m}a_u\tilde n_{um}$.

Solving analytically for $\Phi_m$ and $n_m$ in the non-uniform
equilibrium of an electron hole seems too difficult. The approach
taken here is to perform the required integrations numerically.
 Evaluating $\tilde
n_{um}$ is performed using the same code for each $m$, but using the
different $\omega_m$ in eq.\ (\ref{mweight}). The inner products
we need are $\bra{s}|\tilde V\ket{u}=\int_{-\infty}^\infty s^*(z)\tilde
n_{u}(z) dz$.

\subsection{The external continuum wave contribution}

In the region far outside the hole, $|z/4|\gg 1$ where
$T=\tanh(z/4)\to\pm1$ and $S=\sech(z/4)\to0$ in eq.\ (\ref{ketu}), the
normalized continuum modes (taking them to be antisymmetrically
outward propagating) are
\begin{equation}
  \label{outercont}
 \ket{p}=A\etothe{i\sigma_zpz/4}=\sigma_z{(-15p^4+225p^2-120) +ip (p^{4}-85p^{2}+274)\over
      [8\pi(p^2+1^2)(p^2+2^2)(p^2+3^2)(p^2+4^2)(p^2+5^2)]^{1/2}}\etothe{i\sigma_zpz/4},
\end{equation}
where $\sigma_z={\rm sign}(z)$. Thus they are purely sinusoidal
waves. For large enough $|z|$ the influence of the local hole
potential becomes negligible, and such waves should there be
identified with the normal modes of the uniform background plasma.
The applicable normal mode for the present electrostatic
approximation, in the (usually well justified) cold electrostatic limit,
ignoring ion response, has dispersion relation (including the upper
hybrid waves)
\begin{equation}
  \label{colddisp}
  k_\parallel^2/k_\perp^2={\omega^2(\Omega^2+\omega_{pe}^2-\omega^2)
    \over(\Omega^2-\omega^2)(\omega_{pe}^2-\omega^2)}.
\end{equation}
(In our normalized units $\omega_{pe}=1$.)  Assuming that $\omega$ and
$k_\perp$ are prescribed, there is only one $|k_\parallel|$ that
satisfies this dispersion relation; and corresponding to it, the
continuum eigenmode of $V_a$ has $k_\parallel=\sigma_zp/4$. The mode
number $p$ is taken positive, but $k_\parallel$ is signed. In this
subsection we analyze just this single mode. The approach will be
mathematically justified in the following subsection.

For a single continuum mode, one can immediately calculate the value
of $\Phi_m(z)$ for external positions and \emph{inward}
orbit velocity (e.g.\ negative $v$ on the positive-$z$ side) using
$z'=z(\tau)$ and $\tau=-(z-z')/v$, as
\begin{equation}
  \label{Phiwave}
  \Phi_{m}^{wave}(z) = \int_{\sigma_z\infty}^zA{\rm e}^{i[\sigma_zpz'/4+\omega_m
    (z-z')/v]}dz'/v
  ={A{\rm e}^{i\sigma_zpz/4} \over i(\sigma_zpv/4-\omega_m)}={A{\rm e}^{ik_\parallel z} \over
    i(k_\parallel v-\omega_m)},
\end{equation}
where dropping the infinity limit is justified by positive imaginary
part of $\omega$.  This $\Phi_{m}^{wave}$ for the external 
region applies for all $z$ down to where the eigenmode begins to
deviate significantly, because of non-zero hole potential, from the
external wave; obviously that $z$ is of order a few times 4.

The non-adiabatic distribution perturbation for \emph{outward}
(positive $v$ at positive $z$) orbit velocity, by contrast, is
strongly affected by the rapid variation of potential across the hole
at $z\sim 0$ whose effect is carried into the external region by the
particle orbits. However, for large enough $|z|$ the effects of the
hole and earlier parts of the orbit become negligible because of
dephasing between oscillatory contributions from different
velocities. The dephasing effect attenuates the influence on density
in a distance of several times $z_l\sim 1/\omega_m$, which is finite
but (for $m=0$ at least) typically exceeds the hole extent itself
($\sim 4$) because $\omega$ is small.  Thus the very distant
($|z|\gg z_l$) $\tilde n$ perturbations for inward and outward going
velocities are given approximately by integrating $dv$ the same
$\Phi_m^{wave}$ expression for $z\gg z_l$, but with $v$ of opposite
sign.

The total wave density perturbation can then be considered to be a
constant $\tilde V_m^{wave}$ times $\ket{p}$, where
\begin{equation}
  \label{vmwave}
  \tilde V_m^{wave}= \int_{-\infty}^\infty { b_m(v) \over
    i(k_\parallel v-\omega_m)} dv.
\end{equation}
If $f_{\parallel0}$ is an unshifted Maxwellian, and only $m=0$ is
included, the total wave operator
$\tilde V_w\equiv \sum_m\tilde V_m^{wave}$ is proportional to the
plasma dispersion function $Z$, and in the small $k_\parallel/\omega$
limit becomes $\tilde V_w=1+(k_\parallel/\omega)^2$.  However, that
approximation effectively implements the limit
$\Omega\gg \omega_{pe}$, and gives a dispersion relation
$k_\parallel^2/k_\perp^2 = \omega^2/(\omega_{pe}^2-\omega^2)$, rather
than the full cold plasma expression for finite $\Omega/\omega$: eq.\
(\ref{colddisp}). The full expression arises when the $|m|=1$ terms in
the harmonic sum for the kinetic electrostatic dispersion
relation
\citep[see, e.g.][section 4.4.]{Swanson1989}  are also included, to lowest
order in $\zeta_t^2$. The resulting analytic form
is
\begin{equation}
  \label{Vwanal}
  \tilde V_w= 1 + \left(k_\parallel\over\omega\right)^2 +
  k_\perp^2 {T_\perp/T_\parallel\over\omega^2-\Omega^2}.
\end{equation}

The crucial point is that $(\tilde{V}-\tilde{V}_{w})\ket{q}$ is
localized, tending to zero in the region beyond $z_l$, which means
that overlap integrals of the form
$\bra{p}|\tilde{V}-\tilde{V}_{w}\ket{q}$ exist finite over an infinite
domain. An important secondary feature is that, for external positions
$|z|\ge z_d$ (where the effective hole edge $z_d$ is great enough that
$\phi_0(z_d)=0$), the inward velocity part of $\tilde V_m^{wave}$,
consisting of
\begin{equation}
  \label{inward}
  \tilde V_m^{in}= \int_{0}^\infty { b_m(-\sigma_x|v|) \over
    i(-|k_\parallel v|-\omega_m)}\sigma_x d|v|,
\end{equation}
is \emph{exactly} equal to the actual non-adiabatic density
perturbation that accounts for the hole's presence and the full
eigenmode structure. So the cancellation of the part of
$\tilde V-\tilde V_w$ for inward orbits only is exact
$(\tilde V^{in}-\tilde V_m^{in})\ket{q}=0$. Therefore the non-zero
contribution to external region $z$-integrals of
$(\tilde V-\tilde V_w)\ket{q}$ arises only from outward orbits
$(\tilde V^{out}-\tilde V_m^{out})\ket{q}\not=0$. The necessary
integral of outward orbits, however, must be carried out to an upper limit for
which $|z|\gg z_l\gg z_d$. We shall consider explicitly only parallel
distributions $f_{\parallel0}$ that are symmetric in $v$. In that case
$b_m$ is symmetric, and the overlap integrals can be calculated for a
single sign of $v$ and then doubled to give the total.

\subsection{Reducing the continuum to give the dispersion matrix}

Now we discuss mathematically how the previous considerations allow us
to reduce the continuum contribution to effectively a single mode.
Using $s,u\to q,p$ etc to observe our notation for continuum modes,
and supposing them to be normalized ($\bra{q}\ket{p}=\delta_{qp}$), we
write the continuum part of eq.\ \ref{not4} explicitly as an integral
$a(p)dp$ over an amplitude distribution function $a(p)$ rather than a
sum.  Proceeding with no assumption about the size of cross coupling
of secondary modes, the eigenmode equation inner product with mode
$\bra{s}|$ gives the re-expressed \ref{not4} as
\begin{equation}
  \label{eigengen}
  0=(\lambda_s-\lambda_\perp)\bra{s}\ket{s}a_s+\sum_j\bra{s}|\tilde{V}\ket{j}a_j
+\int\bra{s}|\tilde{V}\ket{p}a(p)dp.
\end{equation}
We apply this equation for the three modes $s=4,2,q$. First for the
continuum mode ($\bra{s}|=\bra{q}|$) with the adiabatic terms moved to
the left hand side:
\begin{equation}
  \label{eigqfirst}
  (-\lambda(q)+\lambda_\perp)a(q)=\sum_j\bra{q}|\tilde{V}\ket{j}a_j
+\int\bra{q}|\tilde V\ket{p} a(p)dp.
\end{equation}
We write the continuum integral using $\bra{q}|\tilde V_w\ket{p}=
\tilde V_w \delta_{qp}$ as 
\begin{equation}
  \label{contintegral}
  \int \bra{q}|\tilde{V}\ket{p}a(p)dp=\int\bra{q}|\tilde{V}_{w}+(\tilde{V}-\tilde{V}_{w})\ket{p}a(p)dp
  =\tilde{V}_{w}a(q) +\int\bra{q}|(\tilde{V}-\tilde{V}_{w})\ket{p}a(p)dp.
\end{equation}
As can be seen in Fig. \ref{modeplots}, the form of the continuum
modes in the inner region is almost independent of $p$. Actually for
small $p$, which is our interest here, the differences are even
smaller than that figure shows. Therefore, to that degree of
approximation, we can replace $\ket{p}$ with $\ket{q}$ in the final
term, giving $\int\bra{q}|(\tilde{V}-\tilde{V}_{w})\ket{p}a(p)dp\simeq
\bra{q}|\tilde{V}-\tilde{V}_{w}\ket{q}\int a(p)dp$, and obtain the
equation
\begin{equation}
  \label{eigq}
  (-\lambda(q)+\lambda_\perp-\tilde V_w)a(q)=\sum_j\bra{q}|\tilde{V}\ket{j}a_j
+\bra{q}|\tilde V-\tilde V_w\ket{q}\int a(p)dp.
\end{equation}
Since the right hand side is a weak function of $q$, this equation
implies that $a(q)$ is a resonant function of $q$ centered on the
$q$-value for which its coefficient on the left hand side is
approximately zero. We can write the coefficient of $a(q)$ using eq.\
(\ref{eigenvalues}) $-\lambda(q)=k_\parallel^2+1$ and eq.\
(\ref{Vwanal}) as
\begin{equation}
  \label{}
  k_\parallel^2+k_\perp^2+1-\tilde
  V_w\simeq k_\perp^2\left(1-{T_\perp/T_\parallel\over\omega^2-\Omega^2}\right)
  +k_\parallel^2\left(1-{1\over\omega^2}\right)
= {1\over4^2}\left({1\over\omega^2}-1\right)\left(q_0^2-q^2\right)
\end{equation}
where
\begin{equation}
  \label{q0def}
  q_0=4k_\perp\sqrt{1+{T_\perp/T_\parallel\over\Omega^2-\omega^2}}
  \Bigg/\sqrt{{1\over\omega^2}-1}.
\end{equation}
The coefficient $(-\lambda(q)+\lambda_\perp-V_w) $ is zero when
$q^2=q_0^2$, which is the dispersion relation of the wave
in a uniform plasma. It is equal to the cold plasma expression
(\ref{colddisp}) when $T_\perp/T_\parallel=1$.

We integrate eq.\ (\ref{eigq})  $dq/(\lambda_\perp-\lambda(q)-V_w)$, recognizing again the approximate independence of $q$ of
the right hand side to find
\begin{equation}
  \label{eigqint}
  \int a(q) dq =\int {dq\over \lambda_\perp-\lambda(q)-V_w}
  \left(\sum_j\bra{q}|\tilde{V}\ket{j}a_j
+\bra{q}|\tilde V-\tilde V_w\ket{q}\int a(p)dp \right).
\end{equation}

In view of the resonant form of the expression for $a(q)$, we can
regard the integral over this resonance $\int a(q) dq$ as quantifying
the total continuum perturbation.  The integral encounters the two
poles of the integrand at $q=\pm q_0$.
% =\pm4k_\perp\omega/\sqrt{1-\omega^2}$.
Near them the real part of $(\lambda_\perp-\lambda(q)-V_w)$ becomes small. Its
imaginary part comes from the small, \emph{necessarily positive},
imaginary part of
$\omega=\omega_r+i\omega_i$. It can then quickly be shown that
$q_0$ has a small positive imaginary part: $q_0=q_{0r}+i\epsilon$.

The infinite integral along the real
$q$-axis can be closed by returning in the upper part of the complex
plane ($\Im(q)$ positive) where the eigenmode tends
to zero. The resonant integral is then just $2\pi i$ times the residue
at the positive $q_0$ pole, which is the mathematical
  justification for our physical assumption of outwardly
  propagating waves. Writing it
$\int {dq\over \lambda_\perp-\lambda(q)-V_w}=1/K$,  we find
\begin{equation}
  \label{Kdef}
  K\simeq{(1/\omega^2-1)q_0\over16\pi i}
  ={k_\perp\over4\pi i}\sqrt{{1\over\omega^2}-1}
\sqrt{1+{T_\perp/T_\parallel\over\Omega^2-\omega^2}}.  
\end{equation}
Substituting elsewhere the
value $q=q_0$ at the integrand's pole, and introducing the shorthand
notation $\int a(p) dp= a_q$, eq.\ (\ref{eigqint}) becomes
\begin{equation}
  \label{eigqint2}
  (K-\bra{q_0}|\tilde V-\tilde V_w\ket{q_0})a_q %\int a(p)dp
  = \sum_j\bra{q_0}|\tilde{V}\ket{j}a_j.
\end{equation}

Treatment of the discrete modes uses again the approximation that
over the resonance $\bra{j}|\tilde V\ket{p}$ can be regarded as
independent of $p$; then for $l=4,2$,
\begin{equation}
  \label{eigenl}
  (\lambda_\perp-\lambda_l)a_l=\sum_j\bra{l}|\tilde{V}\ket{j}a_j
  +\int\bra{l}|\tilde{V}\ket{p}a(p)dp
  =\sum_j\bra{l}|\tilde{V}\ket{j}a_j
  +\bra{l}|\tilde{V}\ket{q_0}\int a(p)dp.
\end{equation}
We regard the amplitudes of the three modes 4,2,$q_0$ as composing a
column 3-vector whose transpose is
\begin{equation}
  \label{avector}
  \a^T=(a_4,a_2,a_q)=(a_4,a_2,\int a(p)dp)
\end{equation}
the three relations in \ref{eigqint2} and \ref{eigenl} can be
considered a matrix equation $\M\a=0$, where
\begin{equation}
  \label{aveq}
  \M=\left[
  \begin{array}{ccc}
    \lambda_4-\lambda_\perp+\bra{4}|\tilde V\ket{4}
    &\bra{4}|\tilde V\ket{2}&\bra{4}|\tilde V\ket{q_0}\\
    \bra{2}|\tilde V\ket{4}
    &\lambda_2-\lambda_\perp+\bra{2}|\tilde
      V\ket{2}&\bra{2}|\tilde V\ket{q_0}\\
    \bra{q_0}|\tilde V\ket{4}
    &\bra{q_0}|\tilde V\ket{2}&-K+\bra{q_0}|\tilde V-\tilde V_w\ket{q_0}
  \end{array}\right]
\iffalse
\left[
  \begin{array}{c}
    a_4\\
    a_2\\
    \int a(p)dp
  \end{array}\right]=0.
\fi.
\end{equation}
The complex determinant of $M$ must be zero for a non-trivial
solution. To connect more directly with the shiftmode calculation,
which simply zeroes the top left coefficient $M_{11}$, and to avoid large values
arising from $M_{33}$,  the scaled quantity (denoted $D$) we actually zero
is the determinant of $M$ divided by the co-factor of $M_{11}$. If
$\omega$ is iterated until $D=0$, the $\omega$ value found will be the
corresponding mode frequency, and the mode structure will be given by
the solution of $\M\a=0$.

\iffalse
Note\footnote{
The ordered perturbative approximations found in prior sections correspond,
in this formulation, to neglecting $\bra{2}|\tilde V\ket{q_0}$,
$\bra{q_0}|\tilde V\ket{2}$ and $\bra{2}|\tilde V\ket{2}$; then
determining $a_2/a_4$ from the second row, and $\int a(p)dp/a_4$ from the third
row.}\fi

Our reduction of the resonant continuum mode uses just the $m=0,\pm 1$
cyclotron harmonics to determine $q_{0}.$ That is justified by
supposing that the finite Larmor radius parameter
$\zeta_t=k_\perp\sqrt{T_\perp/m_e}/\Omega$ is small, and recognizing
$I_m(\zeta_t^2)\propto \zeta_t^{2m}$. If instead $\zeta_t$ is not small (because
$\Omega\not\gg k_\perp v_{t\perp}$), the dependence
$\etothe{-\zeta_t^2}$ predominates in eq.\ (\ref{eq:ftmagnetic}), and
harmonics up to $|m|\sim 3\zeta_t$ approximate a continuous integral
over the perpendicular Maxwellian, with the relevant range of
$\omega_m$ approximately $\omega\pm 3k_\perp v_{t\perp}$. The
resultant frequency spread exceeds $\sim\omega$ for the whistler mode
(for which $\omega < \Omega$) implying that high harmonic
(i.e.\ perpendicular Landau) damping is strong. The continuum mode
contribution itself is then suppressed, and it is reasonable to ignore
coupling to the continuum modes, reducing $\M$ to its upper left
2$\times$2 submatrix.

\section{Evaluation of Matrix Elements}
\subsection{Internal Numerical Evaluation of inner products}

The evaluation of the overlap integrals (forces) that appear in $\M$
for the inner hole region is carried out numerically using methods
that have been documented in detail previously
\citep{Hutchinson2018a}
for mode $\ket{4}$. In summary the process consists, for each relevant
orbit energy $W$, of numerically integrating to obtain the
relationship between $z$ and the prior time $\tau$ for the unperturbed
orbit, and simultaneously accumulating the integral $\Phi_m(z)$ for
each mode, using a discrete (but non-uniform) $z$-mesh.  Trapped
orbits $W<0$ require a sum over all prior bounces in the potential
well, which is represented by a multiplying bounce-resonant
factor. Passing orbits use a single transit across the domain
$-z_d<z<z_d$, and for the discrete modes there is no external
contribution. The continuum mode at the incoming boundary
($z_{in}=-(v/|v|)z_d$), unlike the discrete modes, has a non-zero
value, which is provided by the wave expression
$\Phi_m(z_{in})=\Phi_m^{wave}$, eq.\ (\ref{Phiwave}).  Inside the
prior time integration loop, the overlap integrals
$\int s^*(z)\tilde f_m(z)dz$ are simultaneously
accumulated. Afterwards they are integrated $dv$ over the relevant
range of parallel energy $W$, and summed over relevant harmonics $m$
to give the hole-region contributions to $\bra{s}|\tilde
V\ket{u}$. Except when $s$ is the continuum mode (the bottom row of
$\M$), these internal values provide the full evaluation of the
overlap integrals. For $s=q_0$, however, extra contributions arise
from the external region, which we now describe.

\subsection{External Analytic Integration}

In the external region $|z|>z_d$, the prior integral $\Phi$ giving
$\tilde f$ for discrete modes $\ket{j}$ is non-zero only for outward
moving orbits. We focus for definiteness on $z>z_d$ and positive $v$.
For inward moving orbits, by contrast, $\tilde f$ is zero externally
for the operations $\tilde V\ket{j}$ and
$(\tilde V -\tilde V_w)\ket{q_0}$. Consequently, the external
contribution to the forces $\bra{q_0}|\tilde V \ket{j}$ and
$\bra{q_0}|\tilde V -\tilde V_m\ket{q_0}$ arises from the external
integration $\int_{z_d}^\infty q_0^*(z) \tilde n dz$ only for
\emph{outward} orbits, indicated by superscript $out$ on $\tilde V$.

\begin{figure}
  \centering\includegraphics[width=0.7\hsize]{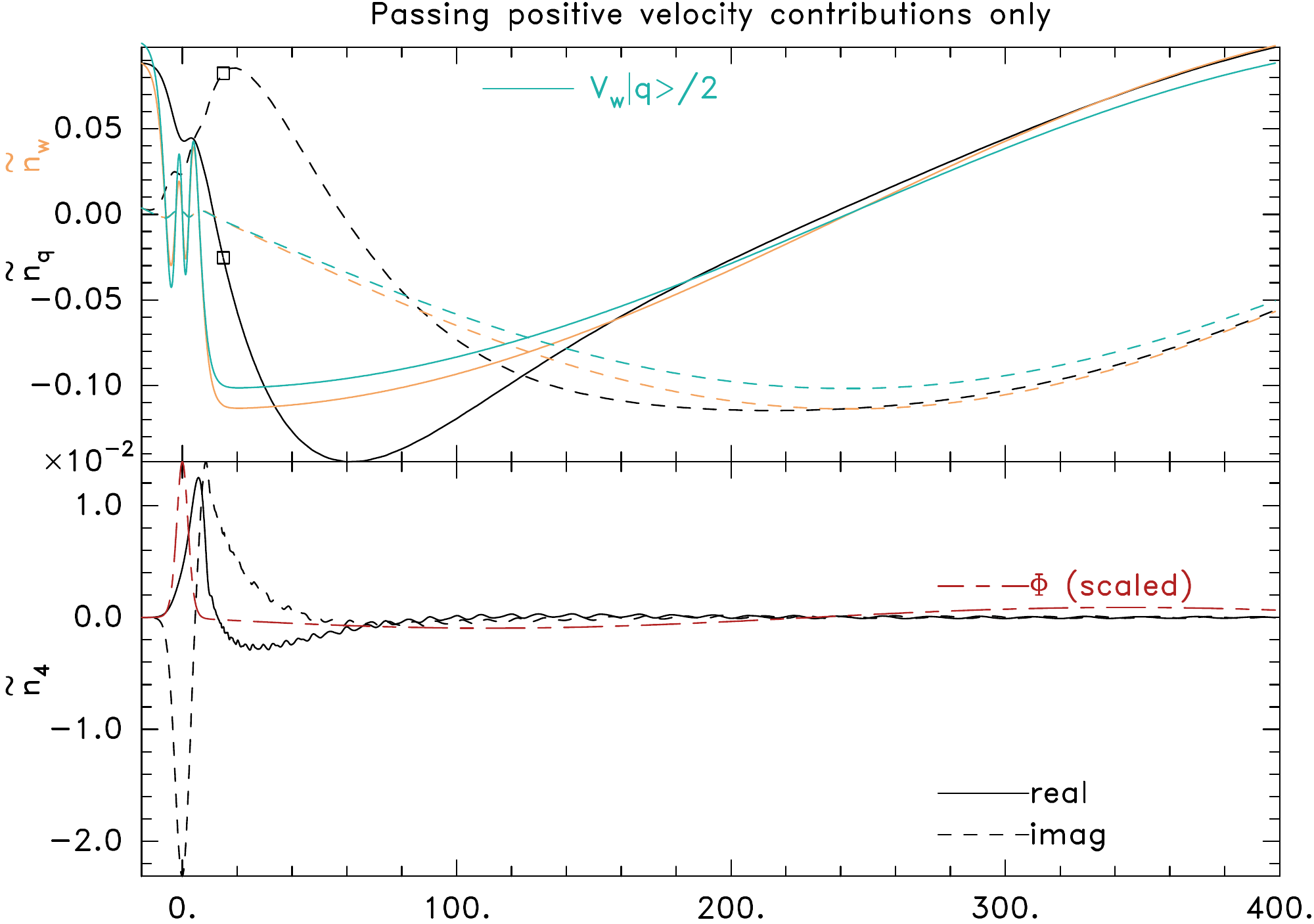}
  \caption{Illustrating the process of calculating the inner products
    involving $\ket{q_0}$ and $\ket{4}$. Passing density contributions
    from positive, outward, velocity orbits integrated over all
    relevant energies. The square point locates the edge of the inner
    hole region, chosen here as $z_d=15$. The scaled $\Phi$ shown is
    for the last included energy $W=6$.\label{externfig2}}
\end{figure}

To assist with understanding, figure \ref{externfig2} shows a case
illustrating the relevant
density perturbations $\tilde n_q=\tilde V^{out}\ket{q}$,
$\tilde n_w=\tilde V_w^{out}\ket{q}$, and $\tilde n_4=\tilde V\ket{4}$
arising from positive velocity orbits. The upper panel shows the
relevant curves for the continuum mode in the inner and outer regions.
The densities $\tilde n_q$ and $\tilde n_w$ are very different in the
inner region $|z|\lesssim z_d$ and well outside it but they converge to each
other in the wave region $|z|\gg z_l$ ($z_l\sim 20$ for this
relatively high frequency illustration). Also shown is the curve of
$\tilde V_w\ket{q}/2$ which is the average of the inward and
outward densities. It has a shape similar to $\tilde n_w$ but is
somewhat smaller because $\tilde V^{in}< \tilde V^{out}$. The
relevant contribution comes from the difference between $\tilde n_q$
and $\tilde n_w$. 

The lower panel shows $\tilde n_4$ which is substantial in the inner
region and converges to zero for $|z|\gg z_l$, on it one can see
residual oscillations caused by discrete contributions at low velocity
approximating the $dv$ integral, which die out at large $z$. In the
upper panel there are also some oscillations but they are barely
visible. The ``$\Phi$ (scaled)'' curve illustrates (only) the shape of
the final contribution to the velocity integral from high velocity,
showing how long the wavelength of oscillations becomes there; its
contribution to $\tilde n$ is small because $f_{\parallel0}$ is
negligible at high $v$.

Now we explain how the external integrations are performed mostly
analytically.  In the external region, the velocity is constant; so
for $\ket{j}$, which has zero perturbed potential there, $\Phi$ is a
simple time delay factor multiplied by its value at the join between internal
and external regions $\Phi(z_d)$:
\begin{equation}
  \label{phi4}
  \Phi_{\ket{j}}(z)=\etothe{i\omega_m(z-z_d)/v}   \Phi_{\ket{j}}(z_d).
\end{equation}
The corresponding expression for the continuum mode, which has
non-zero external potential, may be
found by substituting the wave expression \ref{Phiwave} for  $\ket{q_0}$
which can be integrated analytically to give an extra term, arriving
at
\begin{equation}
  \label{phiq}
  \Phi_{\ket{q_0}}(z)=\etothe{i\omega_m(z-z_d)/v}   \Phi_{\ket{q_0}}(z_d)
  +{A\etothe{ik_\parallel z_d}\over i(k_\parallel v-\omega_m)}[\etothe{ik_\parallel(z-z_d)}
  -\etothe{i\omega_m(z-z_d)/v}].
\end{equation}
Recalling that the wave operator is simply constant, its external
contribution to $\Phi_{\ket{w}}$ for a particular positive velocity is
${A\etothe{ik_\parallel z}\over i(k_\parallel v-\omega_m)}$, which
exactly cancels the first term in the square bracket of
$\Phi_{\ket{q_0}}(z)$. Such cancellation is essential to produce a
finite value for $\bra{q_0}|\tilde V -\tilde V_w\ket{q_0}$. Hence,
when forming
$(\tilde V -\tilde V_w)\ket{q_0}$, the required external prior
integral is
\begin{equation}
  \label{phidiff}
  \Phi_{\ket{q_0-w}}(z)\equiv \Phi_{\ket{q_0}}(z)-\Phi_{\ket{w}}
  =\etothe{i\omega_m(z-z_d)/v}   \Phi_{\ket{q_0}}(z_d)
  %-\Phi_{w}(z_d)]
  -{A\etothe{ik_\parallel z_d}\over i(k_\parallel v-\omega_m)}
  \etothe{i\omega_m(z-z_d)/v}.
\end{equation}
To obtain the total external force we can carry out the inner product
$z$-integration analytically as
\begin{equation}
  \label{extforce}
  \int_{z_d}^\infty q_0^*(z) \Phi_u(z)  dz =
  {A^*\Phi_u(z_d)v\etothe{-ik_\parallel z_d}\over i(k_\parallel  v-\omega_m)}
  + {\delta_{q_0u}|A|^2v\over (k_\parallel v-\omega_m)^2},
\end{equation}
where $\Phi_u$ refers to the ``ket'' mode (j or $q_0-w$), and the
$|A|^2$ term is present only if we are constructing
$\bra{q_0}|\tilde V -\tilde V_w)\ket{q_0}$, as indicated by
$\delta_{q_0u}$. This force quantity is multiplied by the
weighting factor $b_m$ (equation \ref{mweight}) and integrated
(numerically) over parallel velocity so as to produce the total inner
product. For asymmetric $f_{\parallel0}$ the process would need to be
carried out also for negative velocity and $z$, but since we take
$f_{\parallel0}$ to be symmetric the integration is carried out only
for positive $v$ and $z$; the result is then doubled to account for
force exerted at negative $z$. The sum over relevant cyclotron
harmonics $m$ is performed last.

In verification of the fairly complex coding we have compared results
for mode $\ket{4}$ with prior calculations, confirmed certain analytic
self-adjoint properties of the different modes, and debugged the $m=0$
results by benchmark comparisons between two independent
implementations of the calculation (in Fortran and Python).

\section{Results}

\begin{figure}
  \includegraphics[width=.48\hsize]{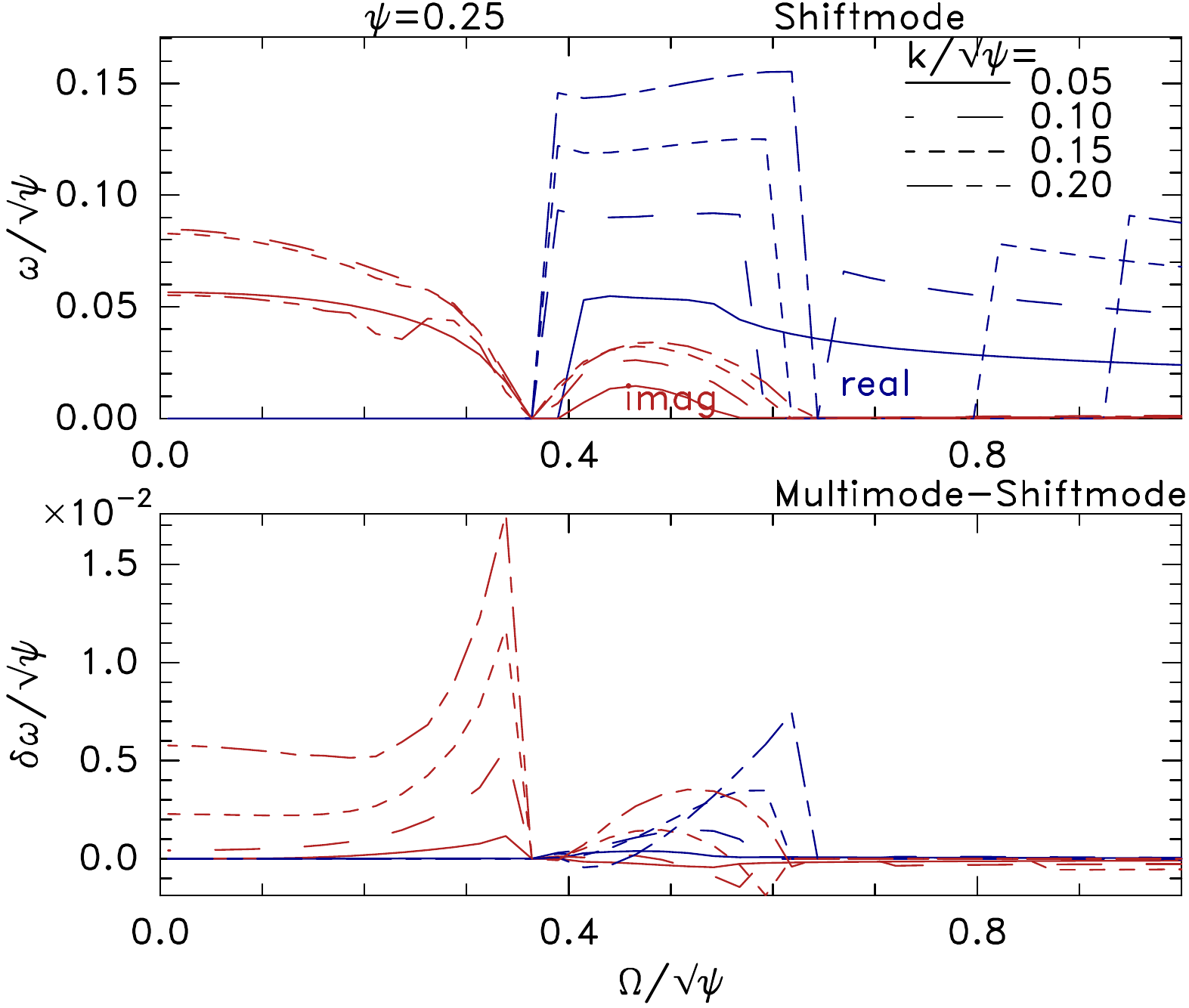}\hskip-2em(a)\quad
  \includegraphics[width=.48\hsize]{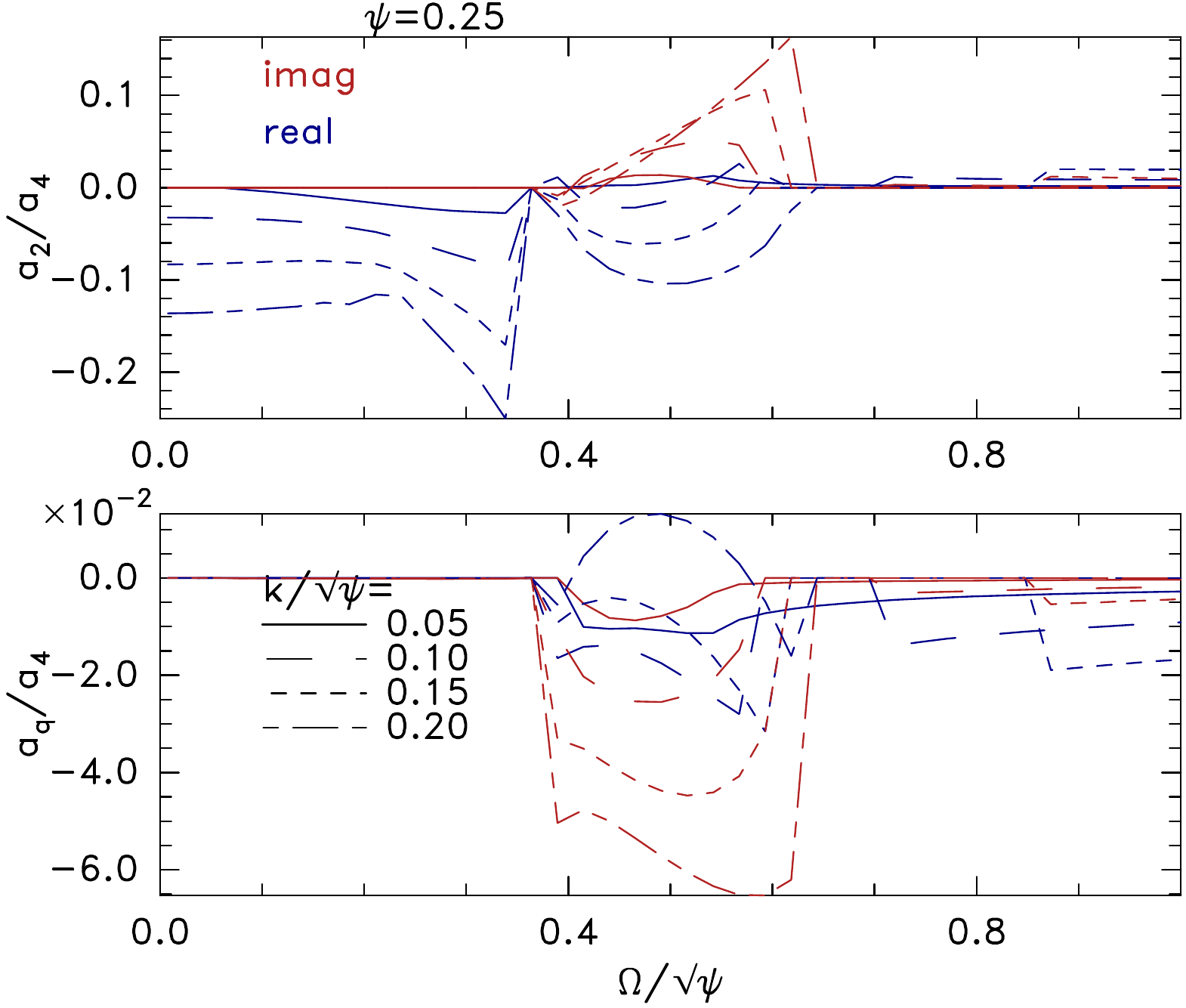}\hskip-2em(b)\ 
  \caption{(a) Frequency $\omega$ and frequency difference between
    multimode and shiftmode calculations $\delta\omega$, and (b)
    additional mode amplitudes $a_2/a_4$ and $a_q/a_4$, all as a
    function of $\Omega$, but for four different values of
    $k_\perp$. \label{ovk25}}
\end{figure}

Fig. \ref{ovk25} gives a comprehensive overview of the the existence
of instability as a function of magnetic field strength, in
appropriately scaled units; frequencies (a) are generally proportional to
$\sqrt{\psi}$ (except at high magnetic field beyond the plotted
range), and the scaled figure changes only a little for different
values of $\psi$.
There are broadly three field strength regimes, as has been documented
in prior publications. At low magnetic field
$\Omega/\sqrt{\psi}\lesssim 0.35$ a purely growing instability exists:
$\omega_r=0$, $\omega_i>0$. At intermediate field,
$0.35\lesssim\Omega/\sqrt{\psi}\lesssim 0.65$ an oscillating
instability $\omega_r>0$ exists with growth rate only of order a
factor of two smaller than the low field regime. At higher field,
$0.65\lesssim\Omega/\sqrt{\psi}$, growth rates $\omega_i$ are either
zero, indicating no instability was found, or extremely small but
positive, which occurs increasingly at higher $\Omega/\sqrt{\psi}$
(and lower $k_\perp$); the unstable high-field cases have real
oscillation frequencies comparable to those in the intermediate
regime. We plot also, in the lower frame of (a) the difference
$\delta \omega$ between the frequencies found by the multimode and
shiftmode calculations. These differences are generally small and the
different $\omega$ scale should be noted. Their imaginary part
$\delta\omega_i$ is always positive in the low field regime, almost
always positive in the intermediate regime, and almost always negative
(and much smaller) in the high-field regime. The real part
$\delta \omega_r$ has some positive differences in the intermediate
regime, especially its upper end, but elsewhere is very small.

The additional mode amplitudes Fig.\ \ref{ovk25}(b) accompanying these
calculations show that the discrete mode 2 real amplitude is
substantial and generally negative in the two lower field regimes, but
slightly positive in the high-field regime. Its imaginary amplitude is
substantial only in the upper part of the intermediate regime. The
continuum mode amplitude is effectively zero in the purely growing
regime, but substantially negative imaginary in the intermediate
regime. It is mostly real, negative, and somewhat smaller in the
high-field regime.

The overall picture is that there are only rather small differences
between the multimode and shiftmode eigenfrequencies, and that almost
all results show the multimode to be somewhat more unstable in the two
lower-field regimes. This summary is in reasonable agreement with
prior expectations. The additional multimode shape flexibility
slightly enhances instability growth rates, but overall the shiftmode
gives rather accurate eigenfrequencies.

%The high-field regime
%observation that the multimode is more stable seems at first more
%counterintuitive.

\begin{figure}
  \includegraphics[width=0.48\hsize]{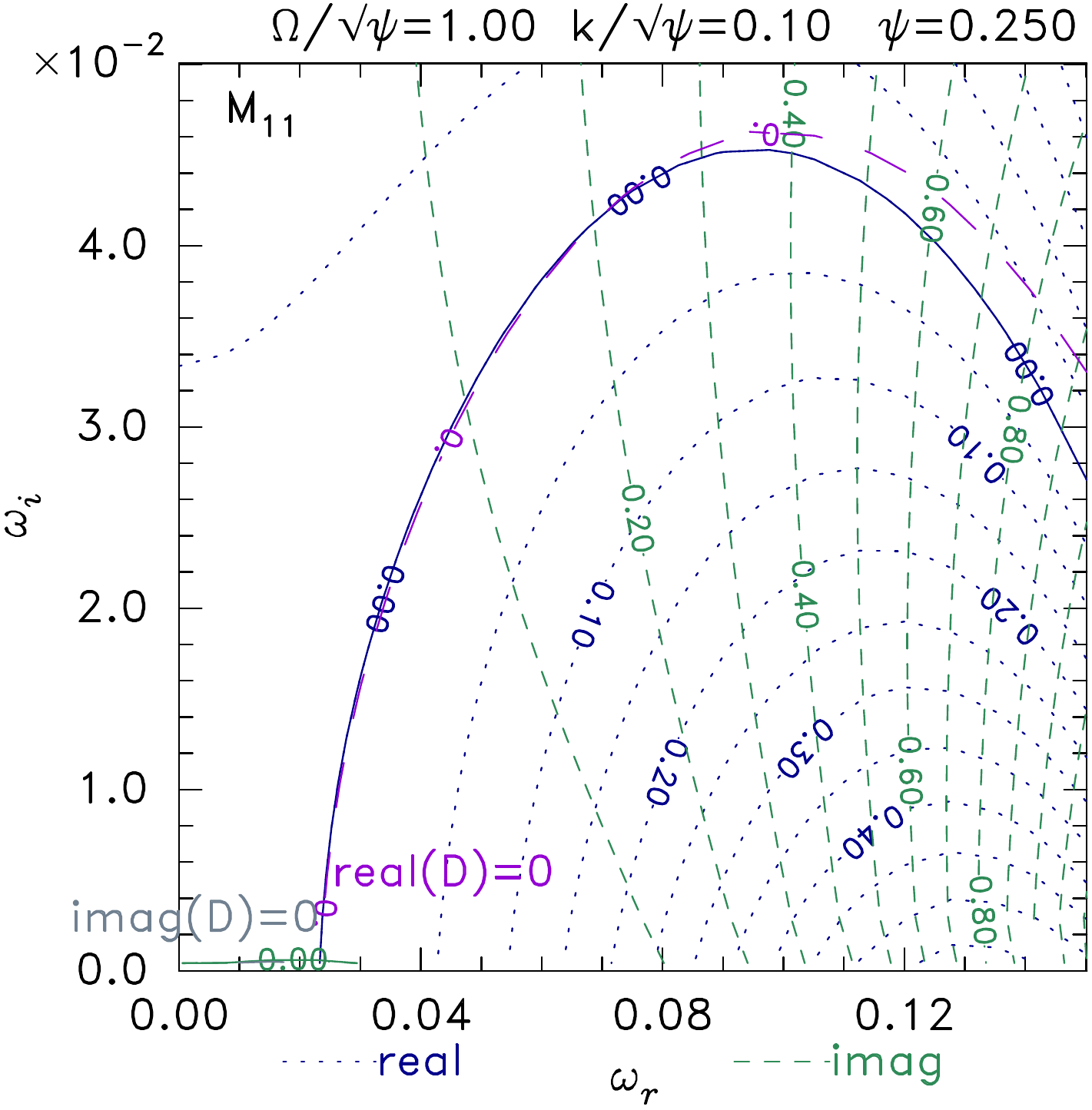}\hskip -2em(a)\quad 
  \includegraphics[width=0.48\hsize]{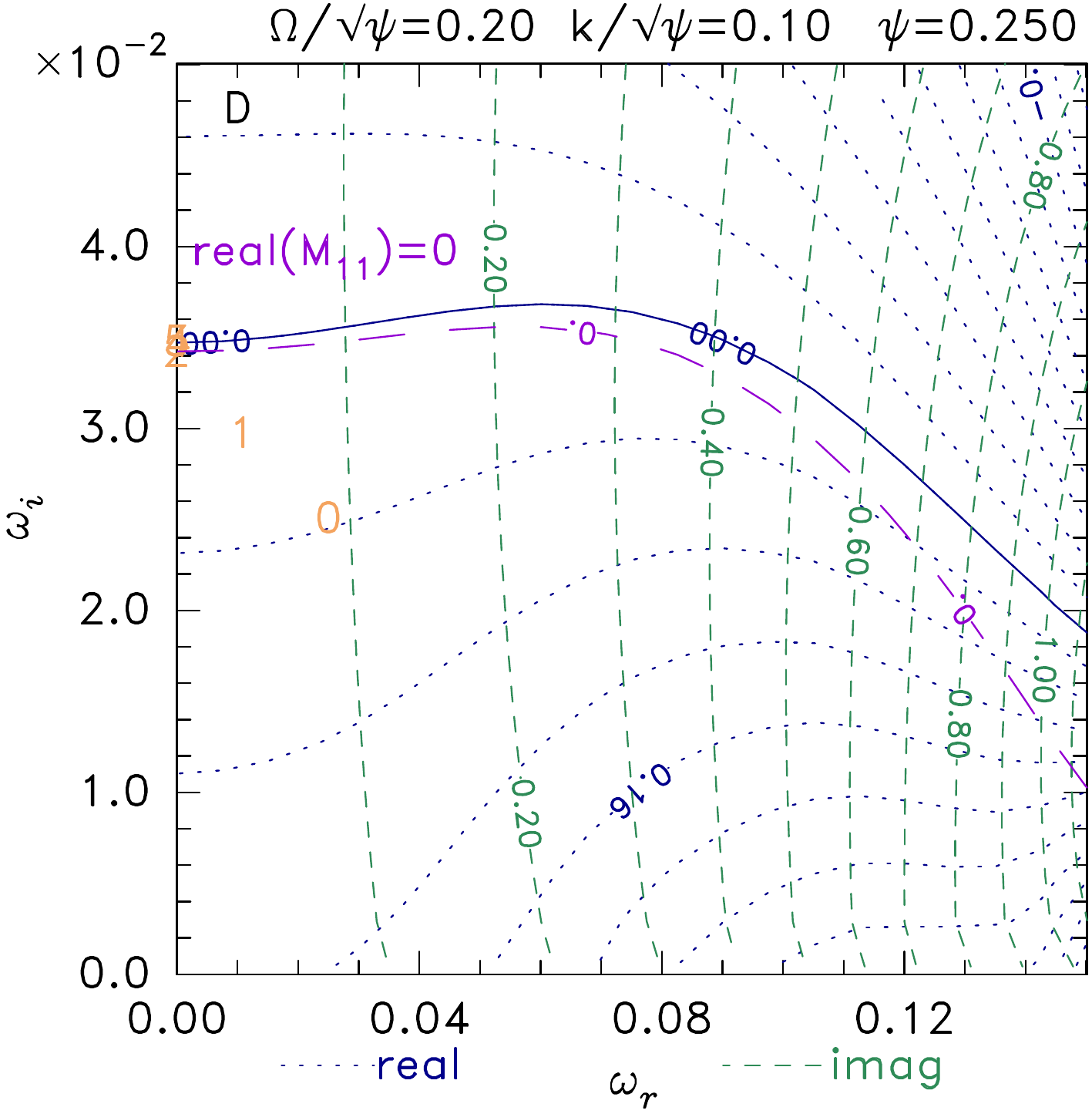}\hskip -2em(b)\ 
  \caption{Contours of force balance quantities which should be zero
    for the eigenfrequency. (a) High magnetic field regime showing
    shiftmode contours and multimode zero contours (long dashed). (b)
    Low field regime showing multimode $D$ contours and shiftmode zero
    contours (long-dashed), but also the Newton steps to
    convergence.\label{hilow}}
\end{figure}

To help explain the mechanisms underlying the three regimes, Fig.\
\ref{hilow} shows contours in the $\omega$ plane of the residual of the
force balance parameter whose roots are found in the prior figures.
The left hand case \ref{hilow}(a) is in the high-field regime, it
plots contours of the real and imaginary parts of the pure shiftmode
force balance $M_{11}$. The eigenfrequency is where the two
zero-contours (blue and green solid lines) intersect, at very small
$\omega_i$. In addition, the zero-contours of $D$, the multimode
equation, are plotted as long dashed lines. Actually the real $D=0$
contour is invisible, but lies just below the (corresponding)
imaginary $M_{11}=0$, where $\omega_i$ is small. The two calculations
agree on the eigenfrequency quite closely; more detail will be given
of this regime later.

The right hand case \ref{hilow}(b) is, by contrast, in the low field
regime in which the instability is purely growing. Contours of $D$ are
shown, plus the zero shiftmode contour of $M_{11}$; the zero
imaginary contours coincide with the vertical axis. The Newton
iterations, which find this intersection precisely, are indicated with
iteration numbers at the corresponding frequencies. The multimode
growth rate slightly exceeds the shiftmode.

\begin{figure}
  \includegraphics[width=0.42\hsize]{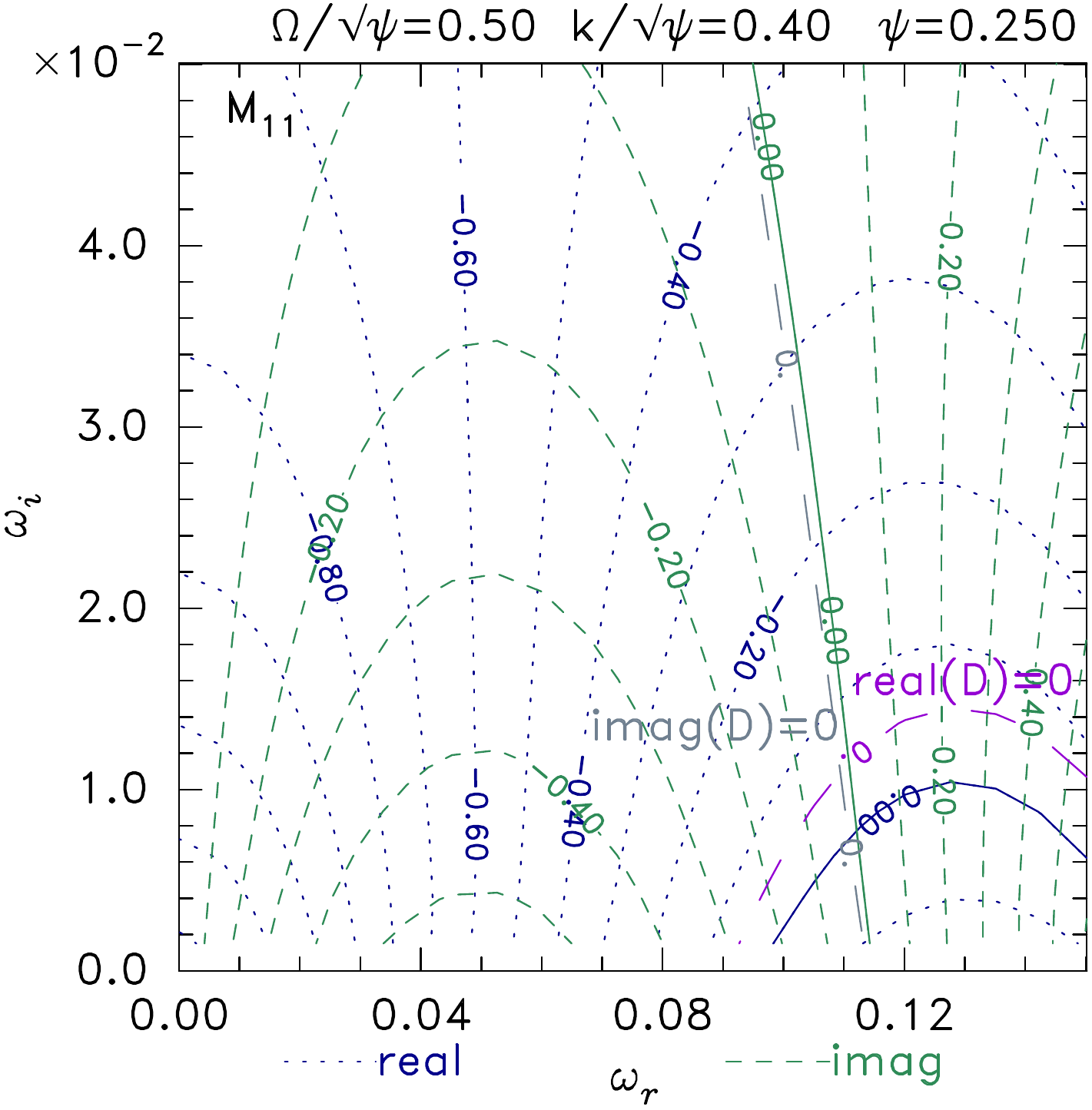}\hskip-2em(a)\quad
  \includegraphics[width=0.56\hsize]{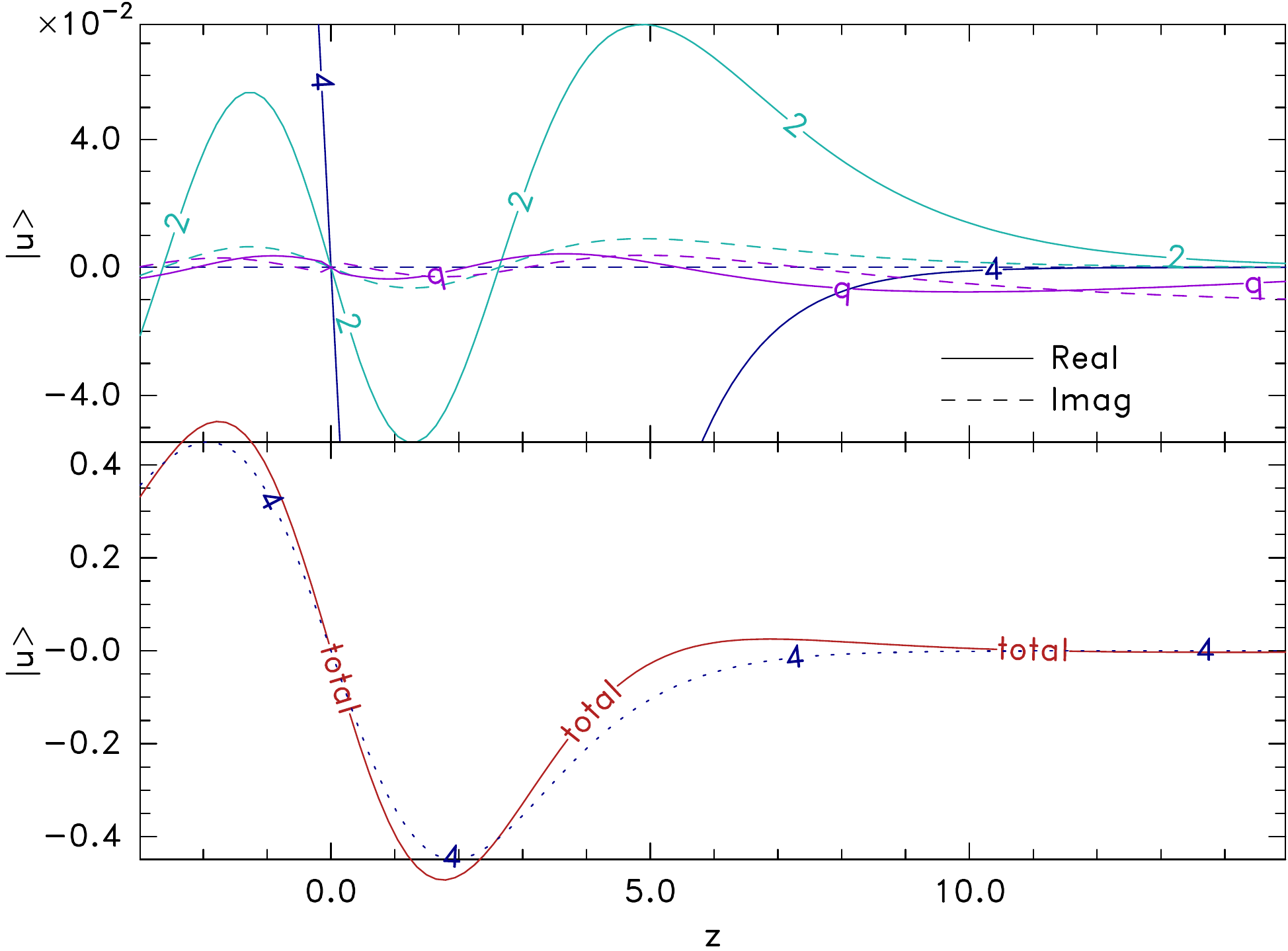}\hskip-2em(b)\ 
  \caption{(a) Intermediate field dispersion contours. (b)
    Contributions to the multimode potential perturbation.\label{intermed}}
\end{figure}

The intermediate field regime is illustrated in Fig.\ \ref{intermed}.
At the left (a) we show the contours, and observe this to be a case
where there is a visible difference between the locations of the
zero-intersections (the eigenfrequencies) in the vicinity of
$\omega/\sqrt{\psi}=0.11+.01i$, for multimode and shiftmode. This
difference arises mostly from the real part of $D$. At the right (b)
are shown plots of the corresponding multimode potential perturbation
as a function of position $z$.  In the upper frame one sees that in
addition to mode 4 contribution, which far exceeds the vertical axis
length, there is a substantial (up to 0.08) real contribution from
mode 2 ($a_2/a_4=-0.21-.025i$). The resultant total (real) mode shape
is shown in the lower panel by the solid line, with the mode 4
contribution alone shown dotted. We observe that the mode 2
contribution increases the total perturbation for $|z|\lesssim0.25$ and
decreases it for $|z|\gtrsim0.25$. The multimode distortion tends to
enhance the influence of deeply trapped particles (concentrated near
$z=0$) and decrease that of marginally trapped particles, which spend
most of their time at large $|z|$. This effect qualitatively explains
the greater growth rate of the multimode instability because, as has
been noted before
\citep{Hutchinson2019}, the deeply trapped particles
drive the instability in this regime, while the marginally trapped
tend to stabilize it.

\begin{figure}
  \center  \includegraphics[width=.5\hsize]{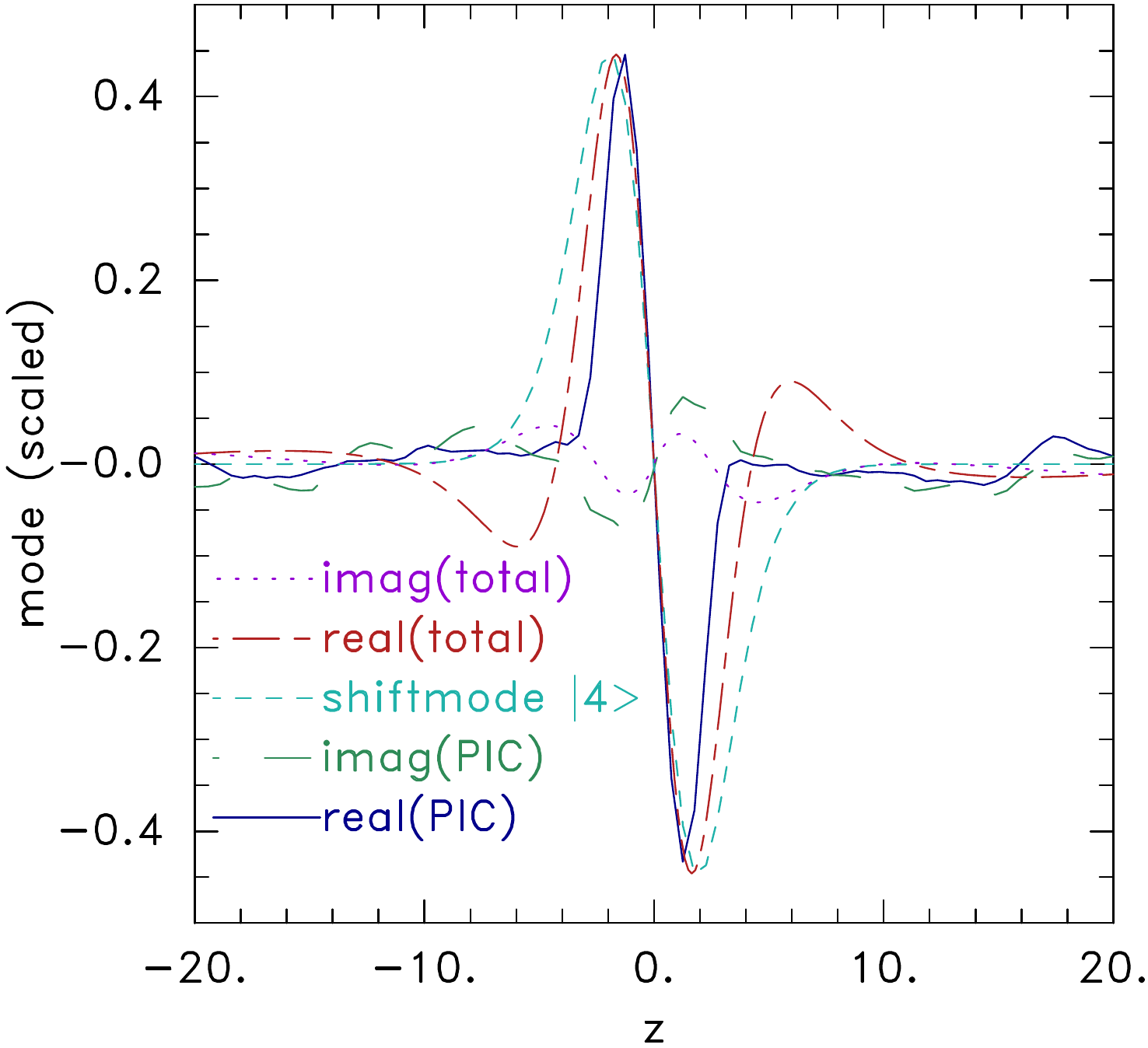}
  \caption{Scaled (total) mode shapes found from multimode analysis,
    compared with shape observed in a PIC simulation having similar
    parameters $\psi=0.16$, $\Omega/\sqrt{\psi}=0.68$,
$k_\perp/\sqrt{\psi}=0.6$.\label{piccomp}}
\end{figure}
In futher support of this interpretation, Fig.\ (\ref{piccomp})
compares the mode shape observed in prior PIC simulations
\citep{Hutchinson2019} with the total mode shape calculated by the
present analysis. The parameters are very similar $\psi=0.16$,
$\Omega/\sqrt{\psi}=0.68$, $k_\perp/\sqrt{\psi}=0.6$, and are at the
upper end of the intermediate field regime. The coupled amplitude of
$\ket{2}$ for the converged eigenfrequency ($\omega=0.12+0.0022i$) is
very substantial: $a_2/a_4=-0.51$. This implies a multimode shape
considerably narrowed from the raw $\ket{4}$ shape, with some
overshoot. The observed PIC simulation mode shape is similarly
narrowed, which is highly suggestive. However, the PIC mode is
slightly narrower and does not show overshoot, which we take as an
indication that for such an extreme case even the multimode treatment
is not quite sufficient to represent the most unstable mode, perhaps
because the expansion in adiabatic eigenmodes is no longer
adequate. We remark that the real parts of the found frequencies agree
very well, but the PIC growth rate was observed to be $\sim 0.004$,
approximately a factor of 2 larger than the multimode calculation. No
analytic pure shiftmode instability was found for these parameters.

\begin{figure}
  \center  \includegraphics[width=.5\hsize]{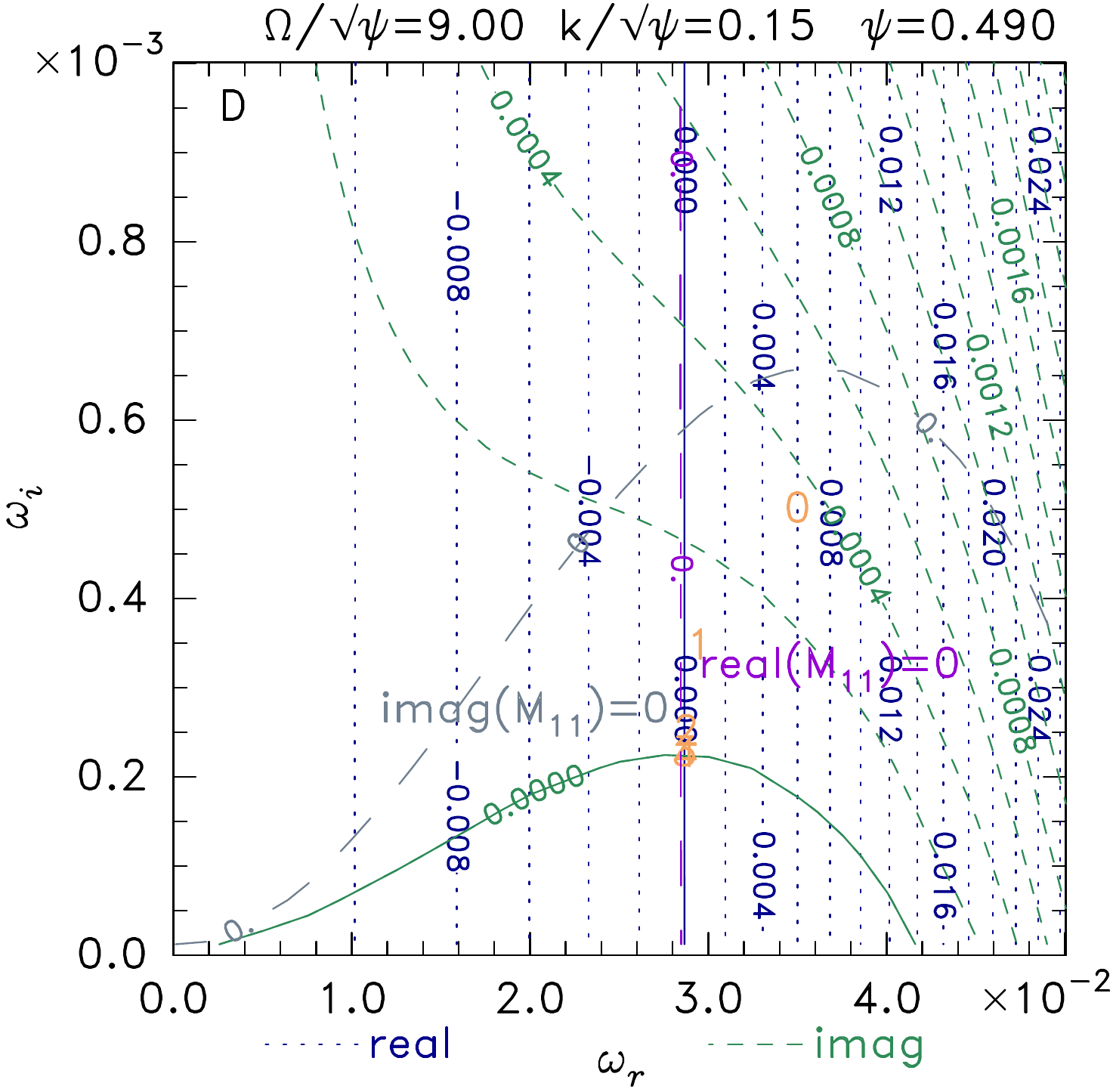}
  \caption{Reduction of the high-field instability growth rate by the
    influence of the continuum mode.\label{hiB}}
\end{figure}

At very high magnetic field $\Omega\gtrsim5$, the approximation of
one-dimensional motion can be used, and dependence on $B$ disappears
from the calculations. It has previously been shown that shiftmode
analysis indicates a very slowly growing overstability that agrees
approximately with simulations
\citep{Hutchinson2019a}.  An example
comparison of contours calculated with multimode and shiftmode
analysis is given in Fig.\ \ref{hiB}.
Multimode analysis finds an unstable eigenfrequency at
$\omega=0.0285+0.0002i$, whose growth rate is nearly a factor of 3
lower than shiftmode.  The differences are essentially entirely caused
by coupling to a small, predominantly real, amplitude of the
continuum, $a_q/a_4=-0.014-0.0018i$. (The negligible influence of
$\ket{2}$, for which $a_2/a_4=0.002-0.0001i$ has been verified e.g.\
by temporarily removing $\ket{q}$ but not $\ket{2}$ from the analysis,
and finding agreement with shiftmode.)  The shiftmode zero-contours
are the long-dashed lines. The real part zero-contour of $\omega$ is
essentially unchanged by inclusion of the continuum mode, but the
imaginary zero-contour is suppressed to lower $\omega_i$. Cases with
much smaller $\psi$ see somewhat less multimode suppression, but they
have a tiny growth rate anyway, since $\omega_i$ scales approximately
as $\psi^{3/2}$. In no case has the high-$B$ instability been
demonstrated to be fully stabilized in this multimode infinite-domain
analysis.

Prior high B-field simulations observed somewhat higher growth rate
than predicted by shiftmode analysis.  The multimode analysis giving
lower growth rate appears to rule out explanation of this observation
by mode distortion, and to favor the cause being suppression in the
simulations of the bounce resonance stabilization by reduced $df/dW$
for shallow trapped orbits. However, the simulations are for finite,
not infinite, parallel domain length.  We can definitely conclude from
the present infinite domain analysis that coupling to the continuum
mode is a significant effect, which is in line with previous
interpretations; but we cannot directly apply our analysis to
simulations of limited domain length.

\section{Discussion}

The analysis here uses a specific potential shape
$\phi_0\propto \sech^4(z/4)$. This is the shape most widely advocated
as ``natural'' for electron holes. It is by no means unique, but it is
part of the family $\sech^l(z/l)$ that satisfies the Debye-shielding
requirement that the potential decays $\propto \exp(-z/\lambda_{De})$
at large $|z|$. Members of this family have the number of discrete
adiabatic eigenmodes equal to $l+1$ (see appendix). Higher $l$
corresponds to a broader profile at its peak and would call for more
than the two discrete antisymmetric modes we have analyzed here. That
possibility also exists for broader profiles of different mathematical
form, not having the convenience of closed analytic expressions for their
eigenmodes. We believe that the current choice is representative, and
that the same qualitative trends would be observed for other profiles.

We also here take the domain boundary to be at infinity. This is the
natural idealization for an isolated hole. However, simulations always
deal with finite (usually periodic) domains. They would require a
different treatment of the continuum modes. External waves would be
restricted to a discrete set of wave-numbers that satisfies the
boundary conditions, and the continuum become discrete. A perhaps more
important difference is that in a finite domain there will be inward
propagating waves (perhaps as components of standing waves) as well as
the outward ones of the present treatment. As a result, the length of
the parallel domain will determine whether certain waves are or are
not coupled resonantly to the perturbed electron hole. The result
could perhaps be stabilizing or destabilizing, depending on domain
size. Accommodating incoming waves would require a major modification of
the analysis, which is beyond our present scope. The infinite domain
analyzed here properly represents truly isolated electron
holes. Exactly periodic analysis would not correctly represent
electron holes in nature, even if they are part of an irregularly
spaced train of holes.

Although our focus here has been on the non-zero $k_\perp$ transverse
instability, some similar adjustments to the stability might arise in
purely one-dimensional oscillatory phenomena. For example, some
regimes of slow electron holes, whose speeds are close to the ion
velocity distribution, experience one-dimensional oscillatory
instabilities \citep{Zhou2017,Hutchinson2021d}, and their growth rates
might be modified in multimode analysis. However, it seems likely that
the multimode corrections would be relatively small, as observed in
our present work. Ion response is negligible for holes moving faster
than a few times the ion sound speed. We have also treated here only
symmetric electron distributions, which limits the quantitative
reliability to holes that are not moving at a large fraction of the
electron thermal speed. The very considerable additional complexity of
including ions and asymmetry hardly seems justified within the present
context. 

\textbf{In summary,} it has been shown how to perform multimode
analysis of electron hole transverse instability, identifying the key
effects as being (1) modification of the effective width of the
eigenstructure, corresponding to an additional discrete eigenmode of
the adiabatic Poisson operator, and (2) coupling to external waves on
the whistler branch, corresponding to a narrow band of the adiabatic
operator's continuum. The width effect slightly increases the growth
rate at low and intermediate magnetic field, and permits relatively
fast growing overstability at magnetic fields up to $\sim20$\% above
the prior shiftmode maximum. The total mode shape is similar to that
observed in PIC simulations. The external wave coupling decreases the
growth rate at high magnetic field.  Despite these adjustments, the
prior single shiftmode analysis is confirmed as giving results very
close, in most cases, to the multimode analysis.

\section*{Acknowledgments}
The work of XC was initially supported by NASA grant NNX16AG82G, and
later in part by US DOE contract DE-SC0019089. No external data was
used. The figures were calculated using code available at
\url{https://github.com/ihutch/multimodeEH}.

%\section*{Appendices}
\appendix
\section{Eigenmode derivation}

The adiabatic operator (generalizing eq.\ \ref{poissonop}) for equilibrium
$\phi_0=\sech^{n-1}(z/l)=S^{n-1}$ is
${d^2\over dz^2}+[n(n+1)S^2-(n-1)^2]/l^2$. Define the operator
for positive $n\in \mathbf{Z}$:
$L_n=l^2\frac{d^2}{dz^2}+{n(n+1)} S^2$; define also the operators
$L^+_n=l\frac{d}{dz}+n T$, $L^-_n=l\frac{d}{dz}-n T$, where
$T=\tanh(z/l)$. Then
\begin{equation}
\begin{split}
  &L^+_n L^-_n=l^2\frac{d^2}{dz^2}-n S^2-n^2 T^2
%  =l^2\frac{d^2}{dz^2}+{(n-1)n} S^2-n^2
  =L_{n-1}-n^2;\\
  &L^-_n L^+_n=l^2\frac{d^2}{dz^2}+n S^2-n^2 T^2
%  =l^2\frac{d^2}{dz^2}+{n(1+n)} S^2-n^2
  =L_n-n^2.\\
\end{split}
\end{equation}
Using these properties, we immediately find that
   \begin{equation}
     L_n L^-_n=(L^-_n L^+_n+n^2)L^-_n=L^-_n (L^+_n
     L^-_n)+n^2L^-_n=L^-_n (L_{n-1}-n^2)+n^2 L^-_n=L^-_n L_{n-1},
     \label{minus}
   \end{equation}
which shows that $L^-_n$ can be considered a raising ladder operator
in the sense that $L_n=L^-_nL_{n-1}(L^-_n)^{-1}$.   
As a result, Eq.(\ref{minus}) leads to
\begin{equation}
  \begin{split}
  L_n L_n ^- L_{n-1}^- ... L_1^-&=L^-_n L_{n-1}L_{n-1}^-
  ... L_1^-=L^-_n L_{n-1}^- L_{n-2}L_{n-2}^- ... L_1^-\\
  & = ...= L_n ^-L_{n-1}^- ... L_1^- L_0
  \end{split}
  \label{nu}
\end{equation}

The eigenfunctions of $L_0$, have the form $\ket{e_0}=\etothe{uz/l}$
with eigenvalue $u^2$, where $u$ can be either real or
imaginary. Specializing to our current case, $n=l+1=5$ in
Eq.(\ref{nu}) we have
\begin{equation}
  L_5 L_5 ^- L_4^- ... L_1^- \ket{e_0}
  = L_5 ^- L_4^- ... L_1^- L_0 \ket{e_0}
  %=L_5 ^- L_4^- ... L_1^-(u^2 \ket{e_0})
  =u^2 L_5 ^- L_4^- ... L_1^- \ket{e_0}.
\end{equation}
Therefore, $\ket{u}=L_5 ^- L_4^- ... L_1^- \ket{e_0}$ is an
eigenfunction of $L_5$ with eigenvalue $u^2$. Since $V_a=L_5/l^2-1=L_5/4^2-1$,
$\ket{u}$ is also an eigenfunction of $V_a$ but the corresponding
eigenvalue becomes $u^2/16-1$. Written out in full,
\begin{equation}
\begin{split}
      \ket{u}=& L_5^- L_4^- L_3^- L_2^-L_1^-\etothe{uz/l} \\
      =&(l\frac{d}{dz}-5T)(l\frac{d}{dz}-4T)(l\frac{d}{dz}-3T)(l\frac{d}{dz}-2T)(l\frac{d}{dz}-T)\etothe{uz/4}\\
      =&\mathrm{e}^{uz/4}\{[-15u^4+(420S^2-225)u^2-945S^4+840S^2-120]T\\
      &\quad\quad\quad+u[u^4+(-105S^2+85)u^2+(945S^4-1155S^2+274)]\},
\end{split}
\label{e5}
\end{equation}
which is exactly Eq.(\ref{ketu}). When the boundary conditions require
$\ket{u}$ to be bounded at $|z| \to \infty$, and $u$ is real, the
polynomial obtained by letting $S\to 0$ and $T\to {\rm sign}(z)$ must
be zero, which restricts the real values of $u$ to that polynomial's
roots. These are the first five positive integers and give the
discrete modes. Imaginary $u$-values give finite $\ket{u}$ at infinity
and are not restricted except by parity considerations: they are the
continuum modes. Evidently the same process can be used to find the
adiabatic eigenmodes of any potential of the form $\sech^{l}(z/l)$
with $0<l=(n-1)\in \mathbf{Z}$. Such a potential has $l+1$ discrete
modes.

\bibliography{JabRef}

\end{document}